\documentclass[preprint]{JHEP3}
\usepackage{amsmath,epsfig}
\usepackage{amssymb,amsfonts}
 \usepackage{latexsym}
  \usepackage{hhline}
   \usepackage{bm}
    \usepackage{pifont}

\renewcommand{\thesection}{\arabic{section}}

\newcommand{\be}{\begin{equation}}\newcommand{\ee}{\end{equation}}%
\newcommand{\bd}{\begin{displaymath}}\newcommand{\ed}{\end{displaymath}}
\newcommand{\bit}{\begin{itemize}}                        
 \newcommand{\eit}{\end{itemize}}                         
\newcommand{\ben}{\begin{enumerate}}                      
 \newcommand{\een}{\end{enumerate}}                       
\newcommand{\baa}{\begin{array}{lll}}                     
 \newcommand{\eaa}{\end{array}}                           
\newcommand{\ba}{\begin{eqnarray}}                        
 \newcommand{\ea}{\end{eqnarray}}                         
\newcommand{\la}{\label}                                  
  \newcommand{\footn}{\footnotesize}                      
\newcommand{\Ds}{\displaystyle}                           
\newcommand{\gev}[1]{\relax\ifmmode{\text{GeV}^{#1}}      
                     \else{GeV$^{#1}${ }}\fi}             
\newcommand{\Gev}{\relax\ifmmode{\text{GeV}}              
                     \else{GeV{ }}\fi}                    
\newcommand{\Mev}{\relax\ifmmode{\text{MeV}}              
                     \else{MeV{ }}\fi}                    
\def\MSbar{\relax\ifmmode\overline                        
            {\rm MS}\else{$\overline{\rm MS}${ }}\fi}     
\def\as{\relax\ifmmode \alpha_s\else{$ \alpha_s${ }}\fi}  
\def\abar{\relax\ifmmode{\bar{a}}\else{$\bar{a}${ }}\fi}  
   \def\etc{\hbox{\it etc.}{ }}
\newlength{\tabcolf} 
\newlength{\tabcols} 
\newlength{\tabcolt} 

\title{Generalization of the BLM\ procedure and its scales \\
in  any order of pQCD.
}
\author{S.~V.~Mikhailov \\
 Joint Institute for Nuclear Research,
Bogoliubov Lab. of Theoretical Physics,\\
141980, Moscow Region, Dubna, Russia\\
\texttt{\email{mikhs@theor.jinr.ru}}
}
\abstract{
The Brodsky--Lepage--Mackenzie procedure is sequentially
and unambiguously extended to any fixed order of perturbative QCD  
beyond the so called ``large--$\beta_0$ approximation''.
As a result of this procedure, the obtained perturbation series 
looks like a continued-fraction representation.
A subsequent generalization of this procedure is developed,
in order to optimize the convergence of the final series, along the 
lines of the Fastest Convergence Prescription. 
This generalized BLM\ procedure is applied to the Adler D function
and also to $R_{e^+e^-}$ in QCD at N$^3$LO.
A further extension of the sequential BLM is presented which makes
use of additional parameters to optimize the convergence of the 
power-series at any fixed order of expansion.  
}

\begin{document}
\section{Introduction}
The first goal of this article is to extend the well-known Brodsky,
Lepage and Mackenzie (BLM) procedure \cite{BLM83} of scale setting to
any fixed order of perturbative QCD (pQCD).
We will show that higher orders of pQCD in the $\overline{\rm MS}$--scheme
unambiguously determine the new scales in the sequentially extended BLM\ prescription.
More specifically, the effects of the coupling renormalization, encoded
in the $\beta$-function coefficients, are absorbed into a set of proper
scales $\mu^2_i$ of the couplings
$a_i\equiv \alpha_s(\mu^2_i)/(4\pi)$
at any fixed order of pQCD.
Thus, in contrast to the prevailing opinion, presented in the 
Particle Data Group booklet ~\cite{PDG2004} 
(see rev. ``Quantum Chromodynamics'' therein),
higher-order corrections can indeed ``fix'' the scales for the
sequentially extended BLM\ (seBLM) procedure.
At the same time, it may happen that this unambiguous seBLM choice of $a_i,~\mu^2_i$ 
is not the ``best'' one from the point of view of the best series convergence, 
so that one should treat these two phenomena separately.
For this reason, our first goal 
includes the analysis of the final seBLM result with an explicit 
demonstration of the above-mentioned problem.
Our second goal following from this problem is to supply the seBLM procedure
with a mechanism resembling the Fast Apparent Convergence (FAC)
\cite{FAC82} (but not coinciding with it) 
in order to improve the convergence of the series.

To simplify the analysis of the structure of the radiative corrections,
renormalization-group invariant quantities, like the Adler function
(D), are considered below.
Quantities like the heavy quark potential $V_{Q}$ \cite{YSchr99}, the
Bjorken sum rules, the Gross-Llewellin Smith sum rule and so on,
see, for example, Ref.\ \cite{GaKar98,BrodskyL95} for a review,
can be considered in the same manner.
Here an empirical relation between the QCD $\beta$--function coefficients
$b_i$, ~$b_i \sim b_0^{i+1}$ has been used that will be explained later.
The hierarchy of the contributions of the coupling renormalization to
the perturbative coefficients at every order of 
$a_s\equiv \alpha_s(\mu^2)/(4\pi)$ is based on this
power law for $b_i$.
This detailed hierarchy
requires a matrix representation for the
perturbation-series expansion (PE) rather than the standard series one.
The mentioned power relation works well at least up to the currently known
coefficient $b_3$ and $N_f=0\div 5$ active flavors 
in contrast to the usually discussed
proposition of the so-called ``large--$b_0$'' limit
that really implies
$-N_f \gg 1$ (~therefore, $b_0 \sim b_1$ in that approximation),
\ see for a review \cite{G03}.
The seBLM procedure is based on this matrix representation;
therefore, it can be formulated in terms of the dynamical characteristics, 
 namely, the $\beta$--function coefficients, rather than in terms of
certain SU$_c$(3)--Casimir operators that may appear at an intermediate stage.
The use of the relation  $b_i \sim b_0^{i+1}$ allows one to go beyond this large--$b_0$ approximation
in a natural way.
Note that this power law should, of course, fail at some higher
order of the PE when its expected factorial divergence starts to become
operative.
The proposed coupling renormalization is worked out (Sec.\
\ref{Sec:the first stage} and \ref{Sec:next stages}) and includes as a
particular case the ``bubble approximation'' elaborated on in
\cite{Neubert95,BeBr95,BaBeBr95}.
The corresponding new perturbation series obtained with the seBLM procedure
is also worked out.
Then the $\beta$--function expansion is performed explicitly for the 4-loop D-function
and, subsequently, the seBLM\ procedure is applied to
this D-function to highlight the advantages and
disadvantages of the procedure in the case of this physical quantity.
Although higher-order expansion coefficients have been calculated for
other observables, mentioned above, the calculation of the beta-function
expansion for these quantities constitutes a different task, 
and asks for a careful treatment.
For this reason, we restrict ourselves here to the consideration of the
$D$ and $R$ functions.

Different kinds of extensions of the BLM\ approach have been discussed
in a number of interesting articles
\cite{Neubert95,BeBr95,BaBeBr95,GK92,BrodskyL95,BGKL96,BEGKS97,HLM02}
which appeared in the last decade.
The important issues about the scheme ambiguity of the BLM\ procedure
\cite{Chyla95}, as well as the role of the anomalous dimensions (for the
corresponding quantities) in the optimization procedures \cite{CKS97},
will not be discussed here, with the corresponding calculations being
performed in the $\overline{\rm MS}$--scheme for massless QCD.
The present approach differs from those works mainly in three items:\\
(i) it is based on a detailed matrix presentation for the PE; \\
(ii) the proposed generalized scheme is formulated explicitly for any fixed
    order of the PE; \\
(iii) all the sources of the coupling renormalization are taken into account
    and absorbed into the coupling scales.

The seBLM\ procedure may not necessarily pertain to an improvement of
the perturbation series.
A relevant example was mentioned even in the pioneering work of Ref.\ \cite{BLM83}.
An improved machinery, which uses the proper scales of the seBLM, applying 
a FAC-inspired optimization, is considered
in Sec.~\ref{Sec:GBLM}. This generalization of the seBLM\ optimizes the perturbation series and
formally completes the developed method in the sense of convergence. 
Because this method employs the $x$-portions of the proper scales of the 
seBLM procedure in order to achieve the best convergence, it is named here
$x$BLM procedure.

This work is outlined as follows. 
In the next section, a convolution representation is proposed,
which can be useful for the interpretation of the BLM\
procedure.
This interpretation naturally leads to a certain kind of the ``skeleton expansion'' and
is presented in a complete form in Appendix \ref{App-Distribution}. 
Appendix \ref{app:A}, referred to in the same section, contains the proof of the convolution representation and the 
necessary formulas for the $\beta$-function coefficients. 
Section \ref{1-loop BLM task} describes in detail the standard BLM procedure
and defines the task, while Sec.\ \ref{Sec:series-matrix} presents the matrix rearrangement 
of the perturbation-series expansion. 
The first stage of the sequential generalization to 
a fixed $N$-order case 
is considered in Sec.\ \ref{Sec:the first stage}.
The remaining steps and the completion of the task are given in Sec.\ \ref{Sec:next stages},
with some technical details presented for the convenience of the reader in App.\ \ref{App-B}.
An application of the developed method to the Adler function is considered in Sec.\ \ref{Sec:D-NLO},
with relevant technical details relegated to App.\ \ref{App-delta}.
Finally, Sec.\ \ref{Sec:GBLM} presents the optimized version of our approach  
that yields a better convergence of the perturbation-series expansion.

\section{Convolution representation for amplitudes}
\la{Sec:Convolution}
Here we rewrite the standard perturbation power-expansion series for an amplitude
in the form of a formal integral representation.
This representation, in contrast to similar ones in \cite{BeBr95,Neubert95}
and in \cite{HLM02}, does not involve integration over the intrinsic momentum $k$.
Even more, it is not related to a Feynman integral over the momenta at all.
The properties of this representation will be discussed in this section
and we shall use it in the next section as a convenient ``perturbative tool''
to interpret the BLM\ procedure by means of the average virtuality flow.

Let us consider the formal perturbation series $s(a_s)$ for the two-point
amplitudes as functions of the external momentum $Q^2$.
The coupling $a_s$
is normalized for the default choice of the scale $\mu^2 = Q^2$.
In this case, the coefficients of the expansion, $d_n$, are numbers in MS--like schemes
due to the cancellation of the logarithms $\ln^k(Q^2/\mu^2 )$, but
they leave behind remnants: the constant parts  ($\ln(C)$) of these logarithms (accompanied by
the $\beta$-function coefficients) are left over in the $d_n$ coefficients and have to
be taken care of.
For further convenience, we introduce a new expansion parameter $A=b_0 a_s$
``normalized'' by the factor $b_0$ so that
\ba \la{eq:seriesA}
&&s(a_s)= d_0 + \sum_{n=1}a_s^n d_n  \equiv S(A)=d_0 + \frac{d_1}{b_0} \cdot \sum_{n=1} A^n D_n
 ~~\mbox{with}~~ D_1=1
  \ea
and define new coefficients $\Ds D_i=\frac{d_i}{d_1 b_0^{i-1}}$
that will simplify the intermediate calculations,
and will help us maintain the contact with the ``large $b_0$'' limit, $b_0\gg1$ at $A\lesssim 1$.
Note that in the real QCD theory, and below the c-quark threshold (for $N_f=3$),
we have at $\mu^2=1~\gev{2}$ the values $b_0=9 \gg 1$ and
$\Ds A(\mu^2)\equiv \alpha_s(\mu^2)\frac{b_0}{4\pi}\approx 0.32 < 1$
at the NLO level.
The running of the coupling $A \to \bar{A}(t)$
(or $a_s \to \bar{a}_s(t)$) follows the renormalization group (RG) equation
\ba
&&    \frac{d}{dt}\bar{A}\equiv B(\bar{A})
=-\left(\bar{A}^2+c_1\bar{A}^3+c_2\bar{A}^4+ \ldots \right) \la{RGalpha}\\
&&   \mbox{with} ~c_i =\frac{b_i}{b_0^{1+i}}, \la{GRcoeff}
\ea
where
$B(A)$ is the modified~$\beta$--function and $t = \ln\left(Q^2/\Lambda^2\right)$
is a natural variable  for MS-like schemes.
At the one loop order we have a well-known solution to (\ref{RGalpha}):
 $\Ds \bar{A}_{(1)}(t) = \frac1{t}$; at the two-loop order the exact solution
$\Ds \bar{A}_{(2)}$ can be realized in terms of the appropriate branch of 
the Lambert function $W_{-1}$, 
see the discussion in Appendix~\ref{app:A},
 and  \cite{Mag99,Mag05} for details.
Beyond the two-loop level, explicit solutions are difficult to obtain.
The loop order $l$ is indicated by the subscript $l$ in parenthesis,
e.g., $A_{(l)}$.

To evaluate the sum in ( \ref{eq:seriesA}), we need an appropriate representation for 
the coefficients $D_n$.
To this end, let us decompose these coefficients into two parts
\be \label{decomp}
D_n = D^+(n) + (-1)^{(n-1)}D^-(n)
\ee
to distinguish the values of the coefficient for even and odd values of the index $n$.
Then, we construct the generating functions ${\cal P}^{+}, ~{\cal P}^{-}$,  
respectively, for these parts  $D^+(n)$, $D^-(n)$ in the following way:
\ba \label{def:P}
D^\pm(n) = \int_0^{\infty}{ \cal P}^{\pm}(\alpha) \alpha^{n-1}
d \alpha 
\ea
with the normalization condition
\be
D_1 = D^+(1) + D^-(1) \equiv ~\int_0^{\infty}\left( {\cal P}^+(\alpha) + {\cal P}^-(\alpha)\right) d\alpha =1.
\ee
At large $\alpha$ the behavior ${\cal P}^{\pm}(\alpha) \sim \alpha^{\gamma +1}e^{-\alpha/c}$
corresponds to the expected asymptotic behavior of the expansion coefficients $D_n$~\cite{LL77} at large $n$ 
\ba
{\cal P}^{\pm}(\alpha) \sim \alpha^{\gamma +1}e^{-\alpha/c} \to D^{\pm}(n)\sim D_n \sim  \Gamma(n+1)~n^\gamma c^{n},
\ea
which is a familiar behavior exhibiting a purely renormalon divergence proportional to $n!$.
Below, we shall construct a representation for the sum (\ref{eq:seriesA}),
based on Eq.\ (\ref{RGalpha}) and on the representation (\ref{def:P}).

\textbf{1-loop integral representation}.
At 1-loop level, evolution (\ref{RGalpha}) leads to a useful
representation for the powers of $\bar{A}_{(1)}$, viz.,
\be
 \label{1-loopRG}
\frac1{n!}\left(- \frac{d}{d t}\right)^n \bar{A}_{(1)} = \left( \bar{A}_{(1)}\right)^{n+1}.
\ee
Substituting (\ref{def:P}) and (\ref{1-loopRG}) into definition (\ref{eq:seriesA})
and \textit{changing} the order of the sum and the integration, we obtain the formal integral
representation
\ba 
S(\bar{A}_{(1)})=d_0 + \frac{d_1}{b_0} \cdot 
\int_0^{\infty}\left\{ {\cal P}^{+}(\alpha)
\left[\exp{\left(-\alpha \frac{d}{dt}\right)} \bar{A}_{(1)}(t)\right] 
+ {\cal P}^{-}(\alpha)
\left[\exp{\left(\alpha \frac{d}{dt}\right)} \bar{A}_{(1)}(t)\right]\right\} d \alpha \ , \nonumber 
\ea
which contains the shift-argument operators. Finally, we find the representation
\ba 
\la{Int-1-loop}
S(\bar{A}_{(1)})= d_0 + \frac{d_1}{b_0} \cdot \int_0^{\infty} \left\{{\cal P}^{+}(\alpha) \bar{A}_{(1)}(t-\alpha) +
{\cal P}^{-}(\alpha) \bar{A}_{(1)}(t+\alpha)\right\}d \alpha 
\ea
that is \textit{linearized } in the coupling $\bar{A}_1$ in terms of the convolution
\be \la{1-convolution}
\int_0^{\infty} \left({\cal P}^{+}(\alpha) \bar{A}_{(1)}(t-\alpha)+ 
{\cal P}^{-}(\alpha) \bar{A}_{(1)}(t +\alpha)\right) d \alpha  \equiv
\langle \bar{A}_{(1)}(t-\alpha) \rangle + \langle \bar{A}_{(1)}(t+\alpha) \rangle.
\ee
Following Neubert's proposal \cite{Neubert95}, it is convenient to consider
the integration in Eqs.\ (\ref{Int-1-loop}) and (\ref{1-convolution}) as an
average of the corresponding coupling $A$ over ${\cal P}$.
This is denoted 
as an average of the coupling $A(t \mp \alpha)$ on 
the right-hand-side (RHS) of (\ref{1-convolution}).

Representation \ (\ref{Int-1-loop}) seems to be close to that
invented by Neubert \cite{Neubert95}, and also by Beneke, Braun, and
Ball \cite{BeBr95,BaBeBr95}, though it turns out to be different.
The main difference between the two representations will be
demonstrated in the next subsection.
Note that representation (\ref{Int-1-loop}) is a ``formal'' one because the change
of the order of summation and integration in its derivation has not been proved.
Nevertheless, the integration over the Taylor expansion of the factors $\bar{A}_{(1)}(t-\alpha)$
and $\bar{A}_{(1)}(t+\alpha)$
in the integrand of Eq.\ (\ref{Int-1-loop}) leads again to the initial series, given by
Eq.\ (\ref{eq:seriesA}), up to any finite order of the expansion.
Therefore, we shall call this Eq.\ (\ref{Int-1-loop}) the ``sum'' of the 
perturbative series (\ref{eq:seriesA}) in the one-loop running-coupling approximation.
The integrand in Eq.\ (\ref{Int-1-loop}) has a pole singularity at $t=\alpha$,
which reflects the fact that this representation is ill-defined.
One can now discuss how the contour should be deformed to give a rigorous sense
to this integral \cite{BB00,BGG04}. The residue of the pole can be taken as
a measure for the uncertainty of the asymptotic series
and it reflects the factorial growth of the perturbation-series coefficients $D_n$
(see \cite{Neubert95,G03,BB00}).

We are not going to discuss here the non-perturbative interpretation of
Eq.\ (\ref{Int-1-loop}). Let us only mention that quantum field models with
an integrable running coupling $\bar{A}_{(1)}(t)$ lead to a finite expression for
the convolution given by Eq.\ (\ref{1-convolution}).
Therefore, the finiteness of the coupling of the analytic perturbation theory
(APT) \cite{SS97,Shir99}, $A^\text{APT}_{(1)}(t) \leq 1$ provides an example for
a convergent perturbation series in the sense of
the convolution representation (i.e., in a weak sense). 
For a brief discussion of this feature
see Appendix \ref{App-Expansion}.

\textbf{$\bm{l}$-loop generalization of the integral representation}.
Let $\bar{A}_{(l)}$ be the solution of the RG equation in the $l$-loop approximation
on the RHS of Eq.\ (\ref{RGalpha}).
Then, the initial series, Eq.\ (\ref{eq:seriesA}),
with the coupling $\bar{A}_{(l)}$ can be represented as
\ba \la{Int-N-loop}
S(\bar{A}_{(l)})=d_0 + \frac{d_1}{b_0} \cdot
\left\{ \int_0^{\infty} {\cal P}^{+}(\alpha)~ \bar{A}_{(l)}\left(t + \alpha \cdot
\frac{\bar{A}^2_{(l)}}{B(\bar{A}_{(l)})}\right)
 d\alpha \, + (\pm \to \mp) \right\}
 \ea
that should be compared with Eq.\ (\ref{Int-1-loop}).  
For example, for the two-loop running of $\bar{A}$, the representation given by 
Eq. (\ref{1-convolution}) for the average coupling, should be replaced by
\ba \la{Int-2-loop}
\langle \bar{A}_{(2)}(t \mp \frac{\alpha }{1+c_1\bar{A}_{(2)}(t)}) \rangle =
\int_0^{\infty} {\cal P}^{\pm}(\alpha)\bar{A}_{(2)}\left(t \mp 
\frac{\alpha}{1+c_1\bar{A}_{(2)}(t)}\right)
 d\alpha \, .
  \ea 
Representation (\ref{Int-N-loop}) can be proved in the same way as Eq.\ (\ref{Int-1-loop})
in the one-loop case (see Appendix A).
Its Taylor expansion in the second term of the argument of
$\bar{A}_{(l)} \left(t \mp \ldots \right)$ generates the PE in Eq.\ (\ref{eq:seriesA})
for  $A =\bar{A}_{(l)}$.
In what follows we shall call $S(\bar{A}_{(l)})$ in Eq.\ (\ref{Int-N-loop}) the ``sum'' of the
perturbative series in the $l$-loop running-coupling approximation.

\section{An Illustration:  1-loop BLM\ procedure}
\la{1-loop BLM task}
Here we consider the standard BLM procedure from the ${\cal P}$-distribution point of view.
First, we rewrite the series, Eq. (\ref{eq:seriesA}), in the form of the average
 \be
 S(\bar{A}_{(1)})=d_0 + \frac{d_1}{b_0} \cdot \langle \bar{A}_{(1)}(t-\alpha) \rangle \, .
\ee
Note here that in the one-loop approximation one can take only one of its parts for simplicity 
and then 
use the linear property of this average.
Expanding the coupling in the average $\langle \bar{A}_{(1)}(t-\alpha) \rangle$
around $t_0$ and taking into account $\bar{A}(t_0)= A_0$, one finds
\be \la{blm1}
\langle \bar{A}\left((t-t_0-\alpha)+t_0 \right)\rangle =
A_0-A_0^2(t-t_0-\langle \alpha \rangle) + A_0^3\left((t-t_0)^2-2(t-t_0)\langle \alpha\rangle
+ \langle\alpha^2\rangle \right)
+ \ldots .
\ee
In order to fix an appropriate scale $t_0$, it is necessary to demand that the NLO
coefficient (at the order $A_0^2$) in the expansion (\ref{blm1}) or 
some particular part of it
(to be defined later) should be nullified.
To this end, let us consider first the structure  of $\langle \alpha \rangle \equiv D_{2}$ in detail.

$D_2$ is proportional to the coefficient $d_2$ of the
initial series (\ref{eq:seriesA}) which contains a term proportional to $ b_0$ appearing
due to the coupling renormalization.
In the discussion to follow, it turns out to be convenient to use the following shorthand notation:
\be \la{BLM-LO}
d_2=d_1\cdot(b^{}_0~e + f) \equiv d_1\cdot\left((b_0)^{1}~d_2[1] + (b_0)^{0}~d_2[0]\right).
\ee
Here and below, the square brackets contain the powers of the $\beta$-function coefficient $b_0$
associated with the expansion coefficient $d_2$. 
The coefficient $ d_2[1]$ originates from the 
diagrams shown in Fig.\ 1(a), (b), 
where the ellipses denote other diagrams with the vertex  and the gluon or quark
field renormalization.
The coefficient $d_2[0]$ stems mainly from the 
diagrams displayed in Fig.\ 1(c), (d), 
this coefficient is not responsible for the coupling renormalization.
The notation $d_2[\ldots]$ in (\ref{BLM-LO}) is particularly 
useful in the multiloop case because it allows the inclusion of the powers of the 
higher $\beta$-function coefficients. 
\FIGURE[!hb]{
  \centerline{\includegraphics[width=0.45\textwidth]{
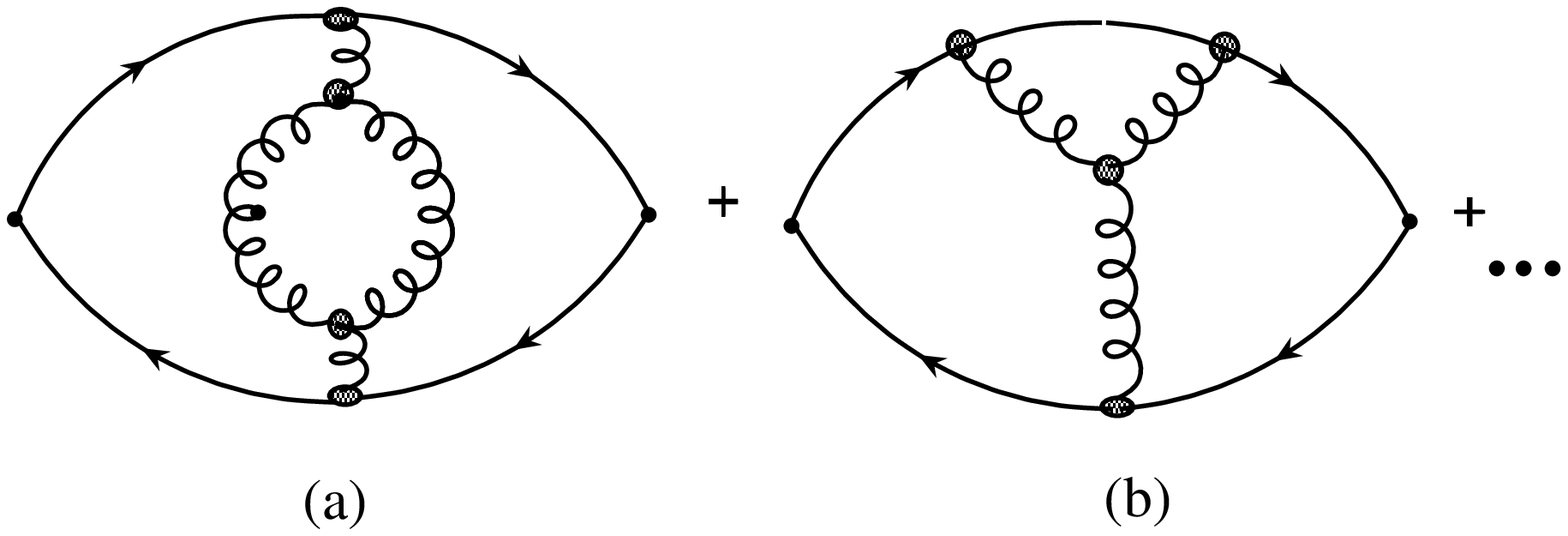}
 ~~~~~~~~~\includegraphics[width=0.45\textwidth]{
 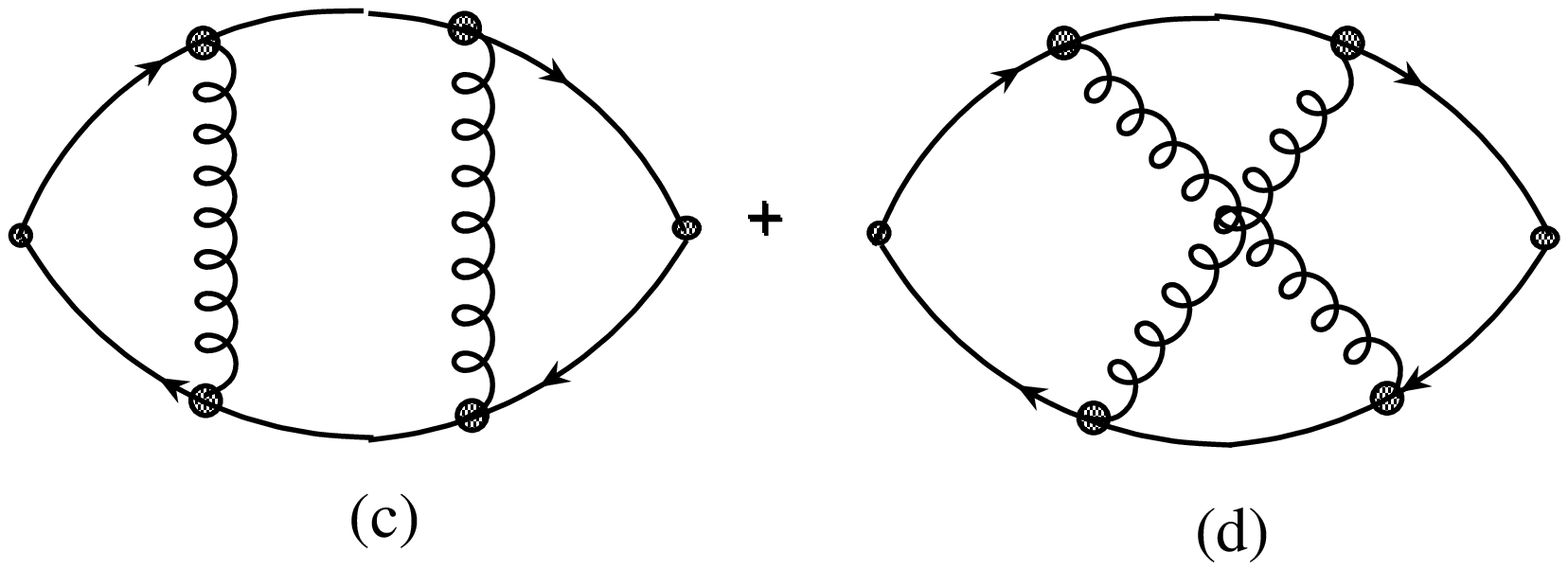}}
   \vspace*{-0.2cm} 
 \noindent
\caption{Diagrams (a,b,$\ldots$) contribute to $b_0~d_2[1]$;
diagrams (c,d) contribute to $d_2[0]$ }
 \label{F:Fig1}}
\noindent Then, we get
\be \la{}\langle \alpha \rangle \equiv D_2 = \frac{d_2}{d_1 b_0}=d_2[1]+ \frac{d_2[0]}{b_0}.
\ee
Two possibilities to fix $t_0$ are considered in the literature:

($i$)
 $t_0=t-D_2\Rightarrow \mu^2=Q^2\exp(-D_2)$ which means that this coefficient is
      nullified in Eq.\ (\ref{blm1}).
This content of the coefficient $D_2$ is included as a whole via the new
expansion parameter $A_0$. This is the meaning of the Fastest Apparent Convergence
(FAC) procedure \cite{FAC82} denoted by
\be \la{FAC-1}
  \langle \bar{A}_{(1)}(t-\alpha) \rangle = A_0 + A_0^2\cdot 0 + O(A_0^3).
\ee
Here the structure of the $O(A_0^3)$ tail looks like
$A_0^3\left(D_3- D_2^2\right)+O(A_0^4)$ \cite{Neubert95}.
The distribution ${\cal P}$ in this approximation can be reduced to $ P_{\text{FAC}}$,
~${\cal P}^{\pm} \to P^{\pm}_{\text{FAC}(1)}=\theta(\pm D_2)~ \delta(\alpha-|D_2|)$.
 
($ii$)
Alternatively, we may demand that
the contribution $d_2[1]$ to the coefficient $D_2$,
which is $\propto b_0$, is nullified so that
$t_0=t-d_2[1]\Rightarrow $ $\mu^2=Q^2\exp(-d_2[1])$.
In this case, only the term $d_2[1]$ responsible for the coupling renormalization
is included by the new expansion parameter $A_0$.
This defines the BLM\ procedure \cite{BLM83,BrodskyL95}
\be \la{BLM-1}
  \langle \bar{A}_{(1)}(t-\alpha) \rangle = A_0 + A_0^2\cdot \frac{d_2[0]}{b_0}
  +O(A_0^3) .
\ee
\noindent The remaining term of this procedure proportional to $A_0^2$ is
suppressed by the inverse power of the ``large $b_0$".
The N$^2$LO term in Eq.\ (\ref{BLM-1}) looks like
$A_0^3\left(D_3- 2D_2 d_2[1] +\left(d_2[1]\right)^2\right)$.
The distribution ${\cal P}$ in the BLM case can be reduced to $P_{\text{BLM}}$,
~${\cal P}^{\pm} \to P^{\pm}_{\text{BLM}(1)}=\theta(\pm d_2[1])~ \delta(\alpha-|d_2[1]|)$.
This determines an intrinsic
``scale'' $d_2[1]$ and shifts the normalization scale from $t$ to $t_0$.

Let us consider now the generating functions ${\cal P}$  from the point
of view of their standard PE.
In this case, ${\cal P}$ can be represented as a formal series; viz.,
\be \la{eq:PE}
{\cal P}^{\pm}(\alpha) = \sum_{n=0} \frac{(\mp 1)^n}{n!} \delta^{\{n\}}(\alpha) \cdot D^{\pm}(n+1)
\ee
that generates just the standard PE, ~$\sum_{n=1} A^n D_n$, in Eq.\ (\ref{eq:seriesA}).
In practice, only a few terms of this series are known. 
Such a broken series in (\ref{eq:seriesA}) corresponds to the broken
series in (\ref{eq:PE}) where every term is strongly concentrated around the
origin $\alpha=0$. However, in the case of a smooth function ${\cal P}$, that is not strongly concentrated
near the origin, (\ref{eq:PE}) provides only an insufficient approximation to
${\cal P}$. 
Indeed,  one should take into account much more terms of the expansion in
Eq.\ (\ref{eq:PE}) or even its infinite subseries to approximate the
real behavior anywhere but not close to the origin, say, near the first extremum of the
${\cal P}(\alpha)$ in $\alpha$. The first two terms
involved into the BLM procedure
$$\delta^{}(\alpha)-\delta^{\{1\}}(\alpha)d_2[1]+ \ldots$$
 generate an infinite model sum, like (at $d_2[1]>0$)
\be \la{eq:PEBLM}
{\cal P}^{+}(\alpha) \to P^+_{{\text{BLM}(1)}}(\alpha)= 
\sum_{n=0}^{\infty} \frac{(-1)^n}{n!} \delta^{\{n\}}(\alpha) \left(d_2[1]\right)^{n}
= \delta(\alpha-d_2[1])
\ee
in contrast to the standard PE in Eq.\ (\ref{eq:PE}) with only a few known terms.
The model distribution $ P_{{\text{BLM}(1)}}^+=$ $\delta(\alpha-d_2[1])$ looks
preferable because this approximation takes into account
at once (admittedly, in a rather crude manner) the main feature of behavior of 
${\cal P}(\alpha)$ near the extremum at $\alpha \approx d_2[1]$.
What parts of the PE in the next higher orders should one involve in the procedure to
improve $P_{\text{BLM}}$?
Or, stated differently, which are the diagram classes that generate these parts?
To clarify this issue (and to define the next approximation of $\cal{P}$), one should analyze
the structure of the $d_n$ coefficients at
N$^n$LO of the PE. This is the task of the next section.

\section{Perturbation expansion -- from series to matrix}
\la{Sec:series-matrix}
In the previous section, we considered the structure of the coefficient $d_2$,
Eq.\ (\ref{BLM-LO}), which is the basis of the standard BLM.
Now, let us extend this representation to the next higher coefficient appealing to the
evident diagrammatic origins of the renormalizations there.
In N$^2$LO, see examples of the diagram elements in Fig. 2(a,b), 
the renormalization of $a_s$ coming from one--gluon exchanges and from
vertices, and quark field renormalization of the same order of expansion generates
contributions proportional to $ a^3 b_0^2,~a^3 b_1$.
These terms should be compared with those originating from Fig.\ 1(a) and Fig.\ 1(b)
in NLO.
A new kind of contribution like $a^3 b_0$ is generated by the two-gluon exchange
with the renormalization of one of these gluon lines/vertices, see Fig.\ 2(c).
The representation for $d_3$ is similar to that in Eq.\ (\ref{BLM-LO}) and
looks like an expansion in a power series in $b_0, b_1,\ldots$, i.e.,
\be
d_3=d_1 \cdot \left( b_0^2 d_3[2,0]+b_1 d_3[0,1]+b_0 d_3[1,0]+d_3[0,0]\right),
\ee
where the first argument $n_0$ of the expansion coefficients $d_n[n_0,n_1,\ldots]$
corresponds to the power of $b_0$, whereas the second one $n_1$ corresponds to the
power of $b_1$, \etc
The coefficient $d_n[0,0]$ represents
so--called~\cite{Chyla95} ``genuine'' corrections
with $n_i=0$ for all possible $b_i$ powers.
One typical diagram generating such contributions is depicted
in Fig.\ 2(e) (compare with the diagrams in Fig.\ 1(c) and Fig.\ 1(d)).
If all the arguments of the coefficient $d_n[\ldots,m,0,\ldots,0]$
to the right of the index $m$ are equal to zero, then we shall omit these arguments
for simplicity and write instead the brief notation $d_n[\ldots,m]$.
\vspace*{-0mm}

\FIGURE[!hb]{
 \centerline{\includegraphics[width=0.95\textwidth]{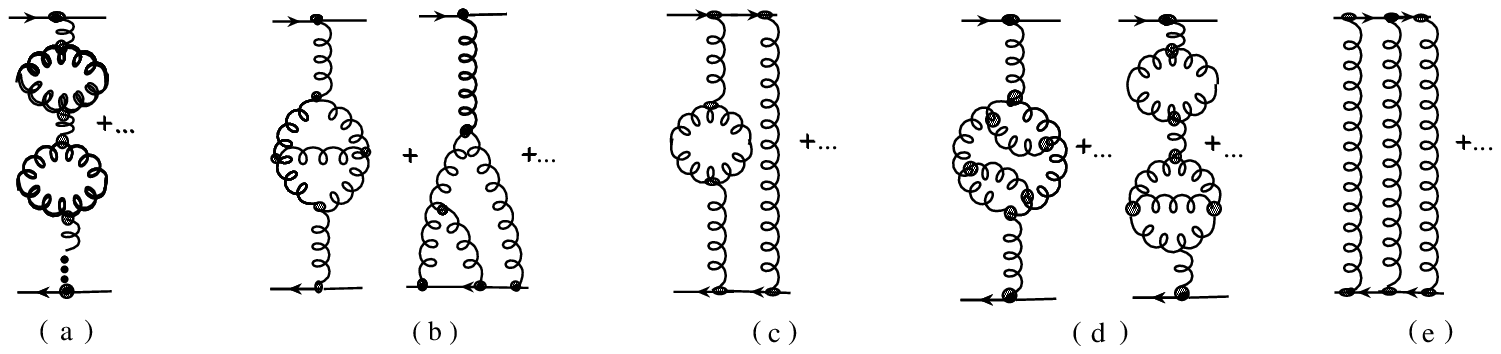}}
 \vspace*{-10mm}
   \caption{
   \footn Diagram elements for $\alpha_s$--radiative corrections contributing, e.g., to the
   photon polarization operator $\Pi$:
(a) a chain of gluon bubbles contributing to the $b_0^2$-term in the coefficient $d_3$.
The ellipses here denote other diagrams with the gluon/vertex renormalization;
(b) diagrams contributing
    to the $b_1$--term in $d_3$ to illustrate the renormalization of gluon fields and vertices;
(c) diagrams that contribute to the $b_0$--term in $d_3$;
(d) diagrams generating $b_2$- and $b_0 b_1$- terms in $d_4$;
(e) diagrams without any intrinsic renormalization contributions to 
$d_3[0]$.
}
 \label{F:Fig2}
 } 
In N$^3$LO the  renormalization of $a_s$ generates contributions
proportional to $ a_s^4 b_0^3,~a_s^4 b_0 b_1,~a_s^4 b_2$ originating
from one--gluon exchanges and/or vertices.
On the other hand, contributions proportional to $a_s^4 b_0^2,~a_s^4 b_1,~a_s^4 b_0$
originate from the mixing of the coupling renormalization stemming from different
gluon lines and/or vertices.
Finally, contributions like $a_s^4$ appear from ``genuine'' corrections, i.e.,
those corrections that come from diagrams which do not contain gluon or quark or
intrinsic vertex renormalizations. Such diagrams do not generate contributions
containing powers of the beta-function coefficients.
Following these lines of argument, the $d_4$ coefficient looks in this brief notation like
\ba \la{d_4}
d_4=d_1 \cdot \left( b_0^3 d_4[3]+b_1 b_0 d_4[1,1]+b_2 d_4[0,0,1]+ b_0^2 d_4[2]+
 b_1 d_4[0,1]+ b_0 d_4[1] +d_4[0]\right).
\ea
The same ordering of the $\beta$-function elements holds
for all higher coefficients $d_n$.
It is convenient for our purposes to present this $\beta$-function structure
in terms of the ``normalized''  variables $\bar{A}$ and $D_n$.
The $D_n$ coefficients have an evident form (in the left column the corresponding
coupling powers $\bar{A}^n$ are shown)
\ba \la{struc-1-step}
\bar{A}^1(t)&&D_1=\underline{1};\nonumber\\
\bar{A}^2(t)&&D_2=\underline{ d_2[1] }+ \frac{1}{b_0}\cdot d_2[0]; \nonumber  \\
\bar{A}^3(t)&&D_3=\underline{d_3[2]}+\underline{c_1 d_3[0,1]}+ \frac{1}{b_0}\cdot\left( d_3[1]+
          \frac{1}{b_0}d_3[0] \right);  \nonumber \\
\bar{A}^4(t)&&D_4=\underline{d_4[3]}+\underline{c_1 d_4[1,1]}+ \underline{c_2d_4[0,0,1]}+  \nonumber \\
&&\hspace{4.8cm}\frac{1}{b_0}\cdot \left(d_4[2]+c_1 d_4[0,1]+
\frac{1}{b_0}\cdot \left(d_4[1]+\frac{1}{b_0} d_4[0] \right)\right);\nonumber \\
\bar{A}^5(t)&&D_5=\underline{d_5[4]}+\underline{c_1 d_5[2,1]}+\underline{c_1^2 d_5[0,2]}+
\underline{c_2 d_5[1,0,1]}
+\underline{c_3 d_5[0,0,0,1]}+ \nonumber \\
&&\hspace{1.cm} \frac{1}{b_0}\cdot \biggl( d_5[3]+c_1 d_5[1,1]+c_2 d_5[0,0,1]+ \nonumber \\
&&\hspace{4.8cm} \left. \frac{1}{b_0}\cdot \left( d_5[2]+c_1 d_5[0,1] +
\frac{1}{b_0} \cdot\left(d_5[1]+\frac{1}{b_0}d_5[0]\right)\right)\right);\nonumber \\
\bar{A}^N(t)  &&D_N= \underline{d_N[N-1]}+ \ldots 
\ea
where $c_i$ are defined in Eq.\ (\ref{GRcoeff}).

Here, we do not discuss how to derive the elements of this representation explicitly for the
known multi-loop results.
It is a separate task 
that is solved for the partial case of the Adler function in Appendix \ref{App-delta}.
Note in this context that similar representations for some physically interesting cases 
were presented in \cite{CKS97,Ch97b}.
In the considerations to follow, we assume that the elements determining the coefficients
$D_n$ in Eq.\ (\ref{struc-1-step})
are known so that they can be used for further considerations.
To clarify the meaning of the structure of the above representation, let us comment on
the nature of its elements.
The first column of the coefficients $ d_n[n-1]$ in Eq.\ (\ref{struc-1-step}) corresponds
to the ``bubble approximation'' that includes the contributions from the diagrams
with the maximum number of bubbles in this order, see Fig.\ 1(a).
These ``bubble'' contributions involved in the extended BLM\ procedure,
were considered before in \cite{Neubert95,BeBr95,BaBeBr95,BrodskyL95}.
However, there are other unsuppressed contributions in Eq.\ (\ref{struc-1-step}),
which have been underlined for a faster identification.
In fact, for the $\overline{\rm MS}$- scheme the values of the coefficients
$c_i$, 
namely,
\be
c_1 \approx 0.79;~~c_2 \approx 0.88;~~c_3\approx 1.9;~~c_4=\mbox{unknown} 
\ee
are of order $1$ where, for definiteness, their estimates have been obtained for $N_f=3$
so that one has no reason to neglect them in Eq.\ (\ref{struc-1-step}).
These terms (cf. the $c_1$-term in $D_3$ in Eq.\ (\ref{struc-1-step}))
originate in part from the diagrams in Fig. 2(b), while
the $c_2$- and $c_1$- terms in $D_4$ in Eq.\ (\ref{struc-1-step}) stem from Fig. 2(d), and so on.
To put our final results on a broader basis,
one may imagine that the uncalculated coefficients $c_i$
are also of the order of unity, $c_i = O(1)$, as we discussed in the Introduction.

We see that the PE series (\ref{eq:seriesA}) subject to the representation (\ref{struc-1-step})
has a complicated structure.
We face two different expansion parameters in Eq. (\ref{struc-1-step}):
first, the coupling $\bar{A}$ for the lines and, second, $b_0^{-1}$ for the columns.
To simplify the treatment of these parameters,
it is convenient to introduce the notation $\bar{A}^i  \cdot y_{ij} \cdot b_0^{-j+1}$
for the contributions in Eq.\ (\ref{struc-1-step}) and $D_i=y_{ij} \cdot b_0^{-j+1}$ 
for their coefficients.
$Y= \{y_{ij}\} $ is a triangular matrix
with the elements $y_{nn} \equiv \mathit{d_{n}[0]}$ on the diagonal.
These diagonal elements 
do not contribute to the coupling renormalization and 
are maximally suppressed by the $b_0^{-1}$ powers in $D_n$,
while the unsuppressed  terms (underlined in Eq.\ (\ref{struc-1-step})) 
are contained in the first column of $Y$
  \ba \la{alph1}
&&y_{11} \equiv \underline{1}; \\
&&y_{21} = \underline{d_2[1]}; \\
&&y_{31}= \underline{d_3[2]+c_1 d_3[0,1]};\la{alph2} \\
&&y_{41}= \underline{d_4[3]+c_1 d_4[1,1]+c_2 d_4[0,0,1]}; \la{alph3} \\
&&\ldots = \ldots~~~~~~~~~~ .\nonumber
\ea
\vspace*{-8mm}

\noindent All these terms originate from the renormalization of a single
coupling/gluon line. As an illustration, see the skeleton diagram in Fig.\ 3(a)
which characterizes all diagrams like those in Fig.\ 2(a),(b),(d).
Note that only the terms $y_{i1}$ survive in the formal ``large--$b_0$'' limit,
$b_0 \gg 1$ at $c_i=O(1)$.

\section{The BLM\ task, first stage of the generalization}
 \label{Sec:the first stage}
Here we construct the first stage of the generalization of the BLM procedure, 
basing it on the hierarchy of the contributions
introduced in the previous section.
We start with the most important unsuppressed  contributions
$y_{i1}$.
The goal is to find a new pair $(t_1,A_{}(t_1))$---following the RG law---which 
is able to nullify all the $y_{i1}$ terms
and then absorb them into the new expansion parameter $A(t_1)$:
\ba \la{Delta1}
\bar{A}_{}(t) \stackrel{RG}{\longrightarrow}  \bar{A}_{}(t_1)\equiv A_1; ~~t_1 &\equiv & t-\Delta_{1}\nonumber \\
\Delta_{1}&\equiv &\Delta_{1}(A_1)=\Delta_{1,0} + A_{1}\cdot \Delta_{1,1}
+ A_{1}^2\cdot \Delta_{1,2} + \ldots
\ea
The shift $\Delta_{1}$ from the original scale $t$ to the first intrinsic scale $t_1$
is going to be presented in the form of a perturbation series in $A_1$
(as first suggested in \cite{GK92}).
The corresponding procedure consists in re-expanding the initial
expression $\bar{A}^i  y_{ij} b_0^{-j+1}$ in terms of the new coupling $A_{1}$ and
subsequent rearrangement of the power series.
This includes the re-expansion of  the coupling $\bar{A}(t)= A(\Delta_1, A_1)$
and its powers 
in terms of $\Delta_{1}$, 
\ba
\bar{A}(t) = A(\Delta_1, A_1)&=& A_1 - B(A_1)\frac{\Delta_1}{1!} + B'(A_1)B(A_1)\frac{\Delta_1^2}{2!}
+ \ldots \nonumber \\
&=& \exp\left(-\Delta_1 B(\bar{A})\partial_{\bar{A}}\right)\bar{A}\mid_{\bar{A}=A_1},
\ea
which can be realized with the operator 
$\exp\left(-\Delta_1 B(A)\partial_{A}\right)[\ldots]\mid_{A=A_1}$,
see, e.g., \cite{APiv02}.
Applying this expansion together with the expansion for $\Delta_1$ in
Eq.\ (\ref{Delta1}) to the initial expression $\bar{A}^i  y_{ij} b_0^{-j+1}$, 
one arrives at the rearranged series in $A_1$, notably,
\ba \label{allloops}
\exp\left(-\Delta_1(A_1) B(\bar{A})\partial_{\bar{A}}\right)\left[ \bar{A}^i  y_{ij} b_0^{-j+1}\right]\mid_{\bar{A}=A_1}
= (A_1)^i~y^{(1)}_{ij} b_0^{-j+1}.  
\ea
The elements of the new matrix $y^{(1)}_{ij}$ 
 are illustrated below in terms of the initial coefficients $D_i$ and read
\ba
\bar{A}^1~\cdot D_1 \to&\left(A_1\right)^1~\cdot& 1;  \\
\bar{A}^2~\cdot D_2 \to&\left(A_1\right)^2~\cdot& D_2 -1\Delta_{1,0}; \la{1-stage-1}\\
\bar{A}^3~\cdot D_3 \to&\left(A_1\right)^3~\cdot& D_3 -2\Delta_{1,0}
\cdot D_2 -\Delta_{1,0}c_1 +\Delta_{1,0}^2-
\Delta_{1,1};
 \la{1-stage-2}\\
\bar{A}^4~\cdot D_4 \to&\left(A_1\right)^4~\cdot&D_4 -3\Delta_{1,0} \cdot D_3+
\left(3\Delta_{1,0}^2-2 c_1 \Delta_{1,0}\right)D_2
-c_2 \Delta_{1,0} +\frac{5}{2} c_1 \Delta_{1,0}^2-
\Delta^3_{1,0}+ \nonumber \\
&& \hspace{4.8cm} \left(2 \Delta_{1,0}-2D_2 -c_1 \right)
\Delta_{1,1} - \Delta_{1,2} ;\la{1-stage-3}\\
\bar{A}^{N}~\cdot D_{N} \to&\left(A_1\right)^{N}~\cdot&D_{N} - (N-1)\Delta_{1,0} \cdot D_{N-1}+  \ldots \nonumber
\ea
The sequential BLM\ procedure requires that the $y^{(1)}_{i1}$
contributions \textit{should cancel}
at each order $A_1^i$ and $i>1$ in the set of Eqs.\
(\ref{1-stage-1}), (\ref{1-stage-2}), (\ref{1-stage-3}),...~.
This requirement completely determines the elements $\Delta_{1,i}$ of the expansion of
$\Delta_{1}$ in the original set of algebraic equations.
Explicit expressions for the first few $\Delta_{1,k}$, and the omitted term in the equations below,
are presented in Appendix \ref{App-B}.
At this point, we compile a few coefficients that are important for our discussion:
\ba
&&\Delta_{1,0} = y_{21} = d_2[1]; \la{Delta-0} \\
&&\Delta_{1,1} = y_{31}-(y_{21})^2 - c_1y_{21} = d_3[2]- d^2_2[1]+
c_1 \left(d_3[0,1] - d_2[1]\right); \la{Delta-1}\\
&&\Delta_{1,2} = y_{41}-3y_{31}y_{21}-2(y_{21})^3 -c_1\cdot \ldots =
d_4[3]-3d_2[1]d_3[2]+2(d_2[1])^3 +
c_1\cdot \ldots
\la{Delta-2}
\ea
Note that the NLO BLM\ correction in Eq.\ (\ref{Delta-1}) cancels in the particular case
$d_3[2] = (d_2[1])^2$, $d_3[0,1]=d_2[1]$.
The first of these terms corresponds to the geometric progression of the leading logarithms
$d^{\text{BLM}}_n[n-1]=\left(d_2[1]\right)^{n-1}$, 
whereas the second one corresponds to the cancellation of sub-leading logarithms.
If one applies these conditions to $\Delta_{1,2}$ in Eq.\ (\ref{Delta-2})
(see also Eq.\ (\ref{B-Delta(1,2)})),
one obtains again the evident ``geometric'' condition $d_4[3] = (d_2[1])^3$
for the cancellation of the leading logarithmic part at the next step
and so on.
This test demonstrates the self-consistency of the calculations.
If the coefficients $d_{n+1}[n]$ in $y_{n1}$ follow the RG law for the
leading logarithms, then they are taken into account already by the first
element $\Delta_{1,0}$, while in order to take into account the sub-leading 
contributions, one needs more subtle conditions.
The absorbtion of such contributions into the proper scales, see, e.g., the $c_1$-terms
in Eq.\ (\ref{Delta-1}), marks the differences between our procedure and other BLM\
extensions, for example, that of Ref.\ \cite{BrodskyL95}.

In this way, one can rearrange step by step for any fixed order $N$ of the PE the
first column $y_{i1}$ in terms of $\Delta_{1}$.
As a result of this procedure, the initial series
can be reduced to a new one that contains only one unsuppressed term
$A_1 \cdot 1$, while
all the others are suppressed by powers of $\Ds b_0^{-1}$, i.e.,
\ba \label{minor-1-step}
\sum_{i\geq j \geq 1}\left(\bar{A}\right)^i y^{}_{ij} \cdot b_0^{-j+1} \stackrel{\text{1 stage}}{\longrightarrow}
\sum_{i\geq j \geq 1}\left(A_1\right)^i y^{(1)}_{ij} \cdot b_0^{-j+1}=
A_1 + A_1 \cdot\sum_{i\geq j \geq 2}\left(A_1\right)^{i-1} y^{(1)}_{i j} \cdot b_0^{-j+1}.
\ea
In  other words, the matrix $Y$ transforms into a new matrix $Y^{(1)}$,
the first column of which is now $y^{(1)}_{i\ 1}=\delta_{i\ 1}$ and
the other few elements are presented in Appendix B, Eq.\ (\ref{final-1-step}).
These changes in the matrix $Y$ (or in the coefficients $D_i$)
are absorbed into the new normalization ``scale'' $t_1=t-\Delta_1$ of the
coupling $A_1=\bar{A}(t_1)$.
Let us stress that no approximation has been used in all steps of the procedure
up to the $A^N$ order.
The above mentioned detailed hierarchy of the contributions is used
only in order to establish the order of the sequential steps, rather than for
performing an approximation.
Indeed, we initiated this rearrangement with the first unsuppressed
$y_{i1}$ column, the most important part of the hierarchy. 
Now we will continue with the second column of $Y^{(1)}$, $y_{i2}^{(1)}$, 
that is suppressed by the power $b_0^{-1}$ 
and proceed in similar manner with this column.

\section{Sequential BLM\ procedure, next stages }
\label{Sec:next stages}
Let us continue with the diagonalization of the matrix $Y^{(1)}$
by rearranging its second column.
First, we single out the $b_0^{-1}$-- suppressed terms on the RHS of Eq.\ (\ref{minor-1-step})
 which are accumulated by its $(Y^{(1)})_{(n-1)}$-- minor part in
\ba \label{n-1minor}
 \sum_{i\geq j \geq 2}\left(A_1\right)^{i-1} y^{(1)}_{ij} b_0^{-j+1} \, .
\ea
The elements of this minor are represented in the right part of Table \ref{struc-2-step}
(put on the right of the double vertical line).
The first power $\Ds b_0^{-1}$-- suppressed terms are the first entries in each row of this column.
Repeating the same procedure as at the first BLM\ stage (see the previous section)
 \begin{table}[th]
\caption{ \label{struc-2-step} 
The structure of $A_1 \cdot \left(A_1\right)^{i-1} y^{(1)}_{ij} b_0^{-j+1}$,
 here $ \tilde{y}_{n2}^{(1)} = y_{n2}^{(1)}/y_{22}$
}
\centerline{
\begin{tabular}{|r|l||c|l|}
\hline
                 &  &                &  \\
$A_1\cdot $&~1~~& ~              & \quad \quad\quad \quad minor elements \\
\hhline{|~|~|=|=|}
            &           &                           &            \\
 $A_1 \cdot$&~0+~~& $\left(A_1\right)^1 \cdot$&~~$\Ds \frac{d_2[0]}{b_0} \cdot 1$ \\
     &      &                 &                          \\
$A_1 \cdot $&~0+~~& $\left(A_1\right)^2 \cdot$&~~$\Ds \frac{d_2[0]}{b_0}\cdot \tilde{y}_{32}^{(1)} +
   \frac{d_3[0]}{b_0^2} $                        \\
     &      &                 &                             \\
$A_1 \cdot $ &~0+~~& $\left(A_1\right)^3 \cdot$&~~$\Ds \frac{d_2[0]}{b_0}\cdot \tilde{y}_{42}^{(1)}+
  \frac{1}{b_0^2}y^{(1)}_{43} +
  \frac{d_4[0]}{b_0^3} $                             \\
     &      &                 &                        \\
$A_1 \cdot $ &~0+~~& $\left(A_1\right)^4 \cdot$&~~$\Ds \frac{d_2[0]}{b_0}\cdot \tilde{y}_{52}^{(1)}+
   \frac{1}{b_0^2}y^{(1)}_{53} +
   \frac{1}{b_0^3}y^{(1)}_{54} +
\frac{d_5[0]}{b_0^4}  $                                                 \\
$\ldots$&~0+~~& $\ldots$&~~$\ldots$ \\
\hline
\end{tabular}
}

\end{table}
with the column $y_{n2}^{(1)}$,
we rearrange again these terms with the help of the
new expansion parameter $A(t_2)$ at the new scale $t_2$ to obtain
\ba \la{Delta2}
\bar{A}_{}(t_1) \stackrel{RG}{\longrightarrow}  \bar{A}_{}(t_2)&\equiv &A_2; \nonumber \\
t_1^{}-t_2 &\equiv &\Delta_{2}=\Delta_{2,0} + A_{2}\cdot \Delta_{2,1}
+ A_{2}^2\cdot \Delta_{2,2} + \ldots~,
\ea
like in Eqs.\ (\ref{Delta1}), (\ref{allloops}). As a result, all the terms from $y^{(1)}_{i2}$, 
except the first one, 
$\Ds y^{(1)}_{22}\equiv d_2[0]$, are transferred into the scale $t_2$ of the new coupling 
$A_2$---therefore, $y^{(1)}_{i 2} \to y^{(2)}_{i 2}=y_{22}\ \delta_{i2}$.
Picking out this diagonal term on the RHS of (\ref{n-1minor}), one arrives at 
the result of the second stage of the procedure  
\ba \label{2stage}
\sum^{n}_{i\geq j \geq 2}\left(A_1\right)^{i-1} y^{(1)}_{ij} b_0^{-j+1}
~\stackrel{\textrm{2~stage}}{\longrightarrow}
~A_2\frac{d_2[0]}{b_0} + 
A_2\sum^{n}_{i\geq j \geq 3}\left(A_2\right)^{i-1} y^{(2)}_{ij} b_0^{-j+1}.
\ea
To fix the new scale $t_2$ from $\Delta_{2,m}$
we use the same procedure as in the previous section.
The first equalities in Eqs.\ (\ref{Delta-0})--(\ref{Delta-2})
remain valid also for the elements $\Delta_{2,m}$
originating from $y_{i2}^{(1)}$ with an evident shifting of all its indices by 1.
Therefore, using Eqs.\ (\ref{final-1-step}) to determine $y^{(1)}_{i2}$ and choosing
the common factor
$\Ds \frac{d_2[0]}{b_0}$ to normalize the elements
($\tilde{y}_{n2}^{(1)} = y_{n2}^{(1)}/y_{22}$),
one arrives at
\ba 
&&\Delta_{2,0} = \tilde{y}^{(1)}_{32} =
\frac{y_{32}}{y_{22}}-2 y_{21} = \frac{d_3[1]}{d_2[0]}-2d_2[1]; \label{2stagedelta} \\
&&\Delta_{2,1} = \tilde{y}^{(1)}_{42} -\left(\tilde{y}^{(1)}_{32} \right)^2
-c_1 \tilde{y}^{(1)}_{32}; \\
&& \ldots \nonumber
\ea
Then substituting Eq.~(\ref{2stage}) into the RHS of Eq.~(\ref{minor-1-step}) 
 one obtains the final result of the second stage
\ba
\sum_{i\geq j \geq 1}\left(\bar{A}\right)^i y^{}_{ij} \cdot b_0^{-j+1}
 \stackrel{\textrm{1.\ stage}}{\longrightarrow}
\ldots
\stackrel{\textrm{2.\ stage}}{\longrightarrow}
A_1\left(1 +  A_2\left(\frac{d_2[0]}{b_0} + 
\sum^{n}_{i\geq j \geq 3}\left(A_2\right)^{i-1} y^{(2)}_{ij} b_0^{-j+1} \right)\right).
\ea
In this way, one can apply a chain of transformations
$Y \to Y^{(1)} \to Y^{(2)} \to \ldots \to Y^{(n-1)}$
from column to column to complete the diagonalization of the matrix $Y$, as it is displayed 
for the first stage in the left column of Table \ref{struc-2-step}.
At each stage of this diagonalization procedure, one will obtain a new coupling $A(t_i)$.
The final result of this successive BLM\ procedure reduces the initial perturbation series,
Eq.\ (\ref{eq:seriesA}), to a series of the special
form\footnote{Let us stress here that the form of this series differs from those
suggested in \cite{BrodskyL95,BGKL96}.}
\ba \la{BLM-final}
&&S(A)\to S_{\text{seBLM}}=d_0 + \frac{ \bar{A}(t_1)}{b_0} \cdot d_1
\Bigg(1 + \frac{\bar{A}(t_2)}{b_0}d_2[0]
\left(1 + \frac{\bar{A}(t_3)}{b_0}d_3[0]
\left(1+\ldots \right) \right) \Bigg)
\ea
that contains only the ``genuine'' elements $d_i[0]=y_{ii}$,
accompanied by the corresponding coupling at its proper scale 
$t_i$, ~$t_i=t - \Delta_{1}-\ldots - \Delta_{i}$.
In other words, the seBLM procedure transforms the standard power 
series $\bar{a}_s^n(t)$  to the series of the products 
$\prod^n_{i=1}\bar{a}_s(t_i)$ keeping invariant the diagonal
elements $y_{ii}$. 
These elements turn out to be the coefficients of the
new series expansion. 

Note that Eq.\ (\ref{BLM-final}) can be easily presented in the form
\be \la{BLM-fraction}
S_{\text{seBLM}} = d_0 +
\frac{a_1 d_1[0]}{\Ds 1-\frac{a_2 d_2[0]}{\Ds 1+ a_2 d_2[0]-
\frac{a_3 d_3[0]}{\Ds 1 + a_3 d_3[0]-\frac{ a_4 d_4[0]}{\Ds \ldots}}}}\, ,
\ee
where $a_i=\bar{a}_s(t_i)$,
a form, motivated by the continued-fraction representation.
The benefit of this representation is that it effects at every step the new coupling
at its new appropriate scale.
This equation itself may become the source of a new approximation for the initial series
$S(A)$.
An instructive interpretation of the procedure from the
``distribution'' point of view
  is presented in Appendix \ref{App-Distribution}.
The diagrammatic sources for both the new expansion coefficients
$d_i[0]$ and the new proper scales $t_i$ are discussed there.
In particular, the skeleton diagram with a single dressed-gluon 
connection (or vertices) in Fig. 3(a) generates the term $\bar{A}(t_1)/b_0 \cdot d_1$
in Eq.\ (\ref{BLM-final}), the shift of the scale $\Delta_1$ is determined 
by the coupling renormalizations appearing from all  corresponding partial
diagrams, see Figs.\ 1,~2(a), and 2(b), and so on. 
Representation (\ref{BLM-final}) corresponds to an
expansion over the classes of skeleton diagrams and provides
one with a particular model for ${\cal P}^{+}(\alpha)$.

 The final result of the seBLM procedure is represented by Eq.\ (\ref{BLM-final})
or Eq.\ (\ref{BLM-fraction}). 
Let us reiterate that the underlying mechanism for obtaining this result is 
due to the rearrangement of the PE via the $Y$-matrix, which also fixes the 
hierarchy of these coupling renormalization contributions.
The constructed procedure takes into account \textit{all} sources of the coupling 
renormalization appearing in the PE and absorbs them completely into the coupling scales, 
just in the spirit of the original BLM method \cite{BLM83}.
At first glance, this looks as an improvement but it is rather formal because 
the seBLM\ procedure
may result in a shifting of the scales into a region where the applicability of pQCD may not be
justified for both $(A(t_i),~t_i)$ and the new expansion coefficients $d_i[0]$.
The question is under which conditions this procedure will provide us 
with a better convergent series.
The standard BLM procedure may improve the NLO approximation, see Eq.\ (\ref{BLM-1}),
if both coefficients $d_2[1]$ and $d_2[0]$ have the same sign.
Similarly, to the above the seBLM procedure may improve the convergence of the series
when the elements of every line of the matrix $Y$ have the same sign.
In this case, the absolute magnitude of the final expansion coefficients
$d_i[0]$ (see (\ref{BLM-final})) is less than that of the initial ones $D_i$.
One, however, should not expect an improvement for the case when the elements in the lines of
$Y$ are alternating in sign and a strong cancellation appears in the coefficients
$D_i=y_{ij} \cdot b_0^{-j+1}$ between different terms.
We shall apply the seBLM procedure to the known N$^3$LO calculation of the $D$--function in
Section \ref{Sec:D-NLO} and argue that this is an example just for the latter case.

\section{\lowercase{se}BLM procedure for the D function}
\la{Sec:D-NLO}
Here we consider  step by step the results of the seBLM procedure
to highlight its advantages and disadvantages from the point of view of the 
convergence of the PE of the Adler D-function. 

The initial well-known series for $D$ \cite{GKL91} can be rewritten
by means of the $\beta$-function coefficients to read
\ba
&&D= 3 \sum_f Q_f^2 \left\{d_0 + d_1 \left[a
  +  a^2 d_2 +  a^3 d_3 +  a^4 d_4 +\ldots \right] \right\},~d_0 =1;~d_1=3C_F;\nonumber\\
&&d_2=b_0 \cdot d_2[1]+ d_2[0];\nonumber\\
&&d_3=b_0^2\cdot d_3[2]+b_1\cdot d_3[0,1]+ b_0\cdot d_3[1]+ d_3[0]; \la{D-NNNLO}\\
&& d_4=b_0^3\cdot d_4[3]+b_0 b_1\cdot d_4[1,1]+
b_2\cdot d_4[0,0,1]+ b_0^2\cdot d_4[2]+ \ldots \nonumber
\ea
However, to recast $D$ into this form is a separate problem which is solved in
Appendix \ref{App-delta} on the basis of the results obtained in \cite{Ch97} 
for the MSSM with light gluinos. 
This new degree of freedom allows us to distinguish the contributions
with different coefficients $b_i$ in $d_3$.
Note that the expressions for the expansion elements in (\ref{D-NNNLO}) remain valid
even including light gluinos that contribute to the $\beta$-function
(Appendix \ref{App-Expansion}).
The explicit expressions for the elements $d_3[m,n]$ are presented in
Appendix \ref{App-delta}, while below their numerical values are displayed and used
\ba
&& d_2 =  b_0\cdot 0.69~+~ \frac1{3};  \la{D-d2}\\
&&\Ds d_3 = b_0^2 \cdot 3.104~~ - ~~b_1 \cdot 1.2~~~ + b_0 \cdot 55.70~-
\left(573.96 +19.83\frac{(\sum_f Q_f)^2}{3(\sum_f Q_f^2)} \right)\, . \la{D-d3numer}
\ea
For the sake of illustration, we substitute the value
$b_0(N_f=3)=9$, $b_1(N_f=3)=64$ into (\ref{D-d3numer}),
\ba
&&\Ds d_3 =~~~~251.1~~-~~~76.8~~~~+~~501.3~~~~~-
~~\left(573.96 + 0 \right) \approx\underline{101.9} ,~~~~~~~~~~~ \la{D-d3numbers}
\ea
in order to compare the contributions stemming from different sources.
Furthermore, we shall apply the seBLM\ procedure to $D$ step by step
to remove, respectively, the $b_0$-contribution to N$^2$LO, the
$b_0^2$ and $b_1$-contributions to N$^3$LO, etc., and the associated results
will be analyzed.

Now we start with our procedure with the original BLM\ scale setting
by transforming the coefficients $d_2,~d_3$
(cf. expressions (\ref{D-d2})-(\ref{D-d3numer})) and the coupling as follows
\ba
 d_2
 &\to &\tilde{d_2} = b_0\cdot 0 + \frac1{3}; \\
\Ds d_3 &\to &\tilde{d_3}= b_0^2 \left(d_3[2]+d_3[0,1]c_1-d^2_2[1]-d_2[1]c_1 \right)+
 b_0(d_3[1]-2 d_2[0]d_2[1]) +
d_3[0] \la{D-d3-1} \\
&& \nonumber \\
&&\phantom{\tilde{d_3}}=b_0^2~(~~~~~2.1555~~~~~~-~~~~1.0251~~~~~~)+b_0(~~55.70 - 0.46~~~~~)+
\ldots\approx \underline{14.7}
\la{D-d3-1-num}  \\
&& \nonumber \\
A(t) & \to & A(t_1); ~t - t_1 =\Delta_{1,0}= d_2[1] \approx 0.69 .
\ea
We see that at the first step
(employing the same condition as when we did with Eq.\ (\ref{D-d3numbers}))
the value of $b_0^2~y_{31}$ is reduced by approximately a factor of 2,
whereas the value of $b_0~y_{32}$ practically does not change.
On the other hand, the value of the total coefficient is
reduced to $\underline{14.7}$ as compared with
the initial value $d_3\approx \underline{101.9}$ in Eq.\ (\ref{D-d3numbers}).
This strong cancellation appears due to the large and negative value of the
``genuine'' term $d_3[0]$. 
The contents of $d_4$ in Eq.\ (\ref{D-NNNLO}) also transform following Eq.\ (\ref{Delta-2}).
Appealing to the results provided in \cite{BChK2002}, which lead to $d_4[3]\approx 2.18$,
one can predict the modification of the``bubble part''
 $d_4[3]$ of the $d_4$,
$$
 d_4[3]\approx 2.18 \to d_4[3]-d_3[2]d_2[1]-2d_2[1](d_3[2]-d_2[1]^2)\approx -3.3\, ,
 $$
which looks rather moderate.

As the next step of seBLM, the modified $\tilde{y}_{31}$ term in Eq.\ (\ref{D-d3-1})
is transferred into $\Delta_{1}$,
following Eqs.\ (\ref{Delta-0}-\ref{Delta-1}),
\ba
\Ds \tilde{d_3} &\to & \tilde{\tilde{d_3}} = b_0^2\cdot 0+
b_1\cdot 0 + b_0 \cdot(d_3[1]-2 d_2[0]d_2[1]) +
d_3[0] \approx -\underline{77} ;\la{D-d3} \\
A(t) & \to & A(t_1); ~t - t_1 =\Delta_{1}= d_2[1]+ A(t_1)\cdot
\left(d_3[2]+d_3[0,1]c_1 - d^2_2[1] - d_2[1]c_1 \right) ;\la{D-t1}\\
&& ~~~~~~~~~~~~~~~~~~~~\Delta_{1} \approx 0.69 + A(t_1)\cdot 1.13 .\la{D-delta-1}
\ea
Notice that one can put $t_1 \approx t-d_2[1]$ for the $A$ argument in Eq.\ (\ref{D-delta-1})
rather than solve Eq.\ (\ref{D-t1}) with respect to $t_1$.
The new absolute value of $d_3$, $d_3 \to \tilde{\tilde{d_3}}\approx -\underline{77}$
is significantly larger than the value of this coefficient at the first step
$\tilde{d}_3\approx \underline{14.7}$.
At the same time, the first perturbation correction to $\Delta_{1}$ in Eq.\ (\ref{D-delta-1})
looks admissible due to the
moderate size of this term. 
Nevertheless, one can conclude that the next step of seBLM\
does not really improve the convergence of the perturbation series due to the large
absolute value of $\tilde{\tilde{d_3}}$.
 
Finally, let us consider the results of the second stage of the seBLM\ procedure.
Following Eq.\ (\ref{2stagedelta}), one finds that
\begin{enumerate}
\item
$t_1-t_2=\Delta_{2,0}=d_3[1]/d_2[0]- 2 d_2[1]\approx 166~(!)$;
therefore, $t_2 \approx t_1-166$ is outside the pQCD domain;
\item this scale-fixing procedure does not lead to a decrease of the $\tilde{\tilde{d_3}}$ term
$\tilde{\tilde{d_3}} \to \tilde{\tilde{\tilde{d_3}}}=d_3[0] \approx -574$
due to a large value of the ``genuine'' term, compared to the contributions
of the other terms in Eq.\ (\ref{D-d3numbers}).
\end{enumerate}
Here we encounter the particularly bad case when the second stage of the seBLM\ procedure cannot be applied 
even to the standard pQCD domain (see the first item above).
But, if one, nevertheless, tries to apply it,
this will not improve the convergence
of the perturbation series (the second item) due to $|d_3[0]| \gg d_3$. 
This happens because the procedure ignores the specific properties of the considered series, like
alternate signs, etc.
Let us recall that a similar effect at the LO was mentioned even in the pioneer work \cite{BLM83}
in considering the branching of the $\Upsilon$--decay, 
$\Gamma(\Upsilon \to \text{hadrons})/\Gamma(\Upsilon \to \mu^+  \mu^-)$.
The next section is devoted to the solution of this problem.
\section{Generalized BLM\ procedure to improve the series convergence}
\la{Sec:GBLM}
\textbf{How to improve the seBLM\ procedure}.
With hindsight, we now would be tempted to think that it would be better
to have not performed the second stage at all and try instead another way to
optimize the value of $\tilde{d}_3$ after the first one.
Indeed, it is not mandatory to absorb the elements $y_{ij}$ 
as a whole into the new scale following BLM, but 
take instead only that part which is appropriate for 
the coupling renormalization.\footnote{
This possibility has already been used for the LO BLM procedure, applied to
the pion form factor 
in \cite{BPSS04}
and termed there $\overline{\text{BLM}}$, see p.13.}
In our case 
it is tempting not to remove the contribution
$y_{31}=d_3[2]+c_1 d_3[0,1]$ completely, as we did in Eq.\ (\ref{D-d3}) 
at the second step above,
but rearrange a part of it and absorb one part into the coupling renormalization
while keeping the other positive part in the remaining expression in order
to compensate the large and negative $d_3[0]$ contribution.
It turns out that it is convenient to introduce the $x$--portion of $y_{31}$, $x~y_{31}$,
in order to absorb it into the scale, see Eq.\ (\ref{Delta-xBLM}),
while its $(1-x)$--portion,
$(1-x)~y_{31}$, is kept explicit in order to be able to cancel the
negative ``genuine'' $d_3[0]$ in (\ref{D-xBLM}).
This trick leads to a modified seBLM\ procedure termed $x$-dependent BLM, ($x$BLM),
\ba
\Ds \tilde{d_3} &\to & \tilde{\tilde{d_3}} =
b_0^2\cdot (1-x)(d_3[2]+c_1 d_3[0,1]) +  b_0 \cdot(d_3[1]-2d_2[0]d_2[1]) +
d_3[0], \la{D-xBLM} \\
t - t_1 &= &\tilde{\Delta}_{1}= d_2[1]+ A(t_1)\cdot
\left(x (d_3[2] + d_3[0,1]c_1) - d^2_2[1]-d_2[1]c_1 \right). \la{Delta-xBLM}
\ea
Let us now establish an ``optimization'' condition, similar to Eq.\ (\ref{FAC-1}),
i.e., put $\tilde{\tilde{d_3}}=0$
to determine the value of $x$.
One has to make sure that in the result the perturbative corrections are improved for both $\tilde{\tilde{d_3}}$
and $\tilde{\Delta}_{1}$
(see the fifth and sixth columns in Table \ref{Tab:DR-3}) in comparison
with those ones in Eqs.\ (\ref{D-d3}) and (\ref{D-delta-1}).
The final result for $D$ is then reduced to
\be \la{D-final}
D= 3 \sum_f Q_f^2 \left\{1 + 3C_F \left[a(\tilde{t}_1)
  + \frac1{3} \cdot a^2(\tilde{t}_1) + 0 \cdot a^3(\tilde{t}_1)\right] \right\},
\ee
As an illustration, the estimates of $\tilde{\Delta}_{1}(Q^2_i)$ 
at $Q^2_1=3~{\rm GeV}^{2}$; $Q^2_2=26~{\rm GeV}^{2}$ are shown 
in the sixth column of Table \ref{Tab:DR-3}. 
\begin{table}[h]
\caption{\la{Tab:DR-3} Results of $x$BLM for
$\tilde{\tilde{d}}_3=0$~ ( $\tilde{\tilde{r}}_3=0$ ) and 
$Q^2_1~~( s_1 )=3~{\rm GeV}^{2}; ~~Q^2_2~~( s_2 )=26~{\rm GeV}^{2}$}
\begin{tabular}{|c|c||c|r|c|c||l|c|c|c|}
\hline
     &                          & &       &                        &         &     &    & &\\
$N_f$&$b_0$& & $x$   & $t-\tilde{t}_1=\tilde{\Delta}_{1}$&$\tilde{\Delta}_{1}(Q^2)$ &   & 
$x$ &$t(s)-t(\tilde{s}_1)=\tilde{\Delta}_{1}$  &$\tilde{\Delta}_{1}(s)$    \\
\hhline{|--|~|---|~|--|-|}
  3  &9        &         & $0.56$& $d_2[1]+ a(\tilde{t}_1) b_0 \cdot 0.18$ &   & 
               & $1.84$   &$d_2[1]-a(\tilde{s}_1) b_0 \cdot 3.1$ & \\
     &         &          &      &   & $ \tilde{\Delta}_{1}(Q^2_1)$  &     &   & &
     $ \tilde{\Delta}_{1}(s_1)$\\
  4   & $\Ds \frac{25}{3}$&$\tilde{\tilde{d}}_3=0$& $0.24$ &$d_2[1]-a(\tilde{t}_1) b_0 \cdot 0.45$&
  $\approx0.58$, 
  &
               $\tilde{\tilde{r}}_3=0$& $2.56$     &$d_2[1]-a(\tilde{s}_1) b_0 \cdot 3.7$ &$\approx -0.27$\\
               &                   &        &       &   &$\tilde{\Delta}_{1}(Q^2_2)$   &  &    &  
               &$\tilde{\Delta}_{1}(s_2)$\\
  5            & $\Ds \frac{23}{3}$&        &$-0.11$&$d_2[1]-a(\tilde{t}_1) b_0 \cdot 1.19$&$\approx0.52$&
               & $3.63$     &$d_2[1]-a(\tilde{s}_1) b_0 \cdot 4.48$ &$\approx -0.05$  \\
\hline
\end{tabular}
\end{table}

It is instructive to apply a similar procedure also to
the observable quantity $\Ds R(s)=\frac{\sigma(e^+e^- \to h)}{\sigma(e^+e^- \to \mu^+\mu^-)}$
associated with $D$:
\ba \la{R-final}
R(s)=D(s)-d_1 \frac{\pi^2}{3} \cdot b_0^2 \bar{a}^3=3 \sum_f Q_f^2 \left\{1 + 3C_F \left[a
  + r_2 a^2 + r_3 a^3\right] \right\},  
\ea
where $\Ds r_1=d_1, ~r_2=d_2, ~r_3=d_3 - \frac{\pi^2}{3} \cdot b_0^2$ ~see, e.g., \cite{GKL91},
$t(s)=\ln(s/\Lambda^2)$ and $a=\bar{a}(t(s))$.
The large and negative $\pi^2$-term, arising due to the analytic continuation,
makes $r_3$ also negative (cf.\ with Eq.\ (\ref{D-d3numer})).
As a result of the $x$BLM procedure, one should replace $ r_2$ by $\tilde{d}_2$ 
and $r_3$ by $\tilde{r}_3$, whereas
the $x$-dependent term in $\tilde{r}_3$ transforms to
$\Ds b_0^2\cdot (1-x)(d_3[2]+c_1 d_3[0,1] - \pi^2/3)$.
Then, to obtain a positive compensating term to cancel $d_3[0]$ at the next step
of the transformation (that leads to $\tilde{\tilde{r}}_3=0$),
one should take $x >1$,
consult the eighth and ninth columns in Table \ref{Tab:DR-3}.
The result of this procedure is exemplified by showing 
partial values of shifts $\tilde{\Delta}_{1}(s_{1,2})$ 
that are placed in the last column of this table.
For these examples, $\tilde{\Delta}_{1}< 0$ and, 
therefore the optimal scale $\tilde{s}_1$ becomes larger than the initial one $s$.

Having developed the $x$BLM\ procedure, we are in the position to
gain control over the size of the coefficients $d_{i+2}$ and the magnitude of the
associated scales $t_i$ to balance their contributions.
This possibility is of paramount importance because we need to control 
\textit{several different} perturbation expansions in the generalized BLM procedure: 
one is the size of the expansion coefficients $d_{i+2}$ like $d_3$
 in  Eq.\ (\ref{D-xBLM}),
the others are for the scale shifts $\Delta_{i}$ like $\Delta_{1}$ in Eq.\ (\ref{Delta-xBLM}).
Finally,
in the result  of this procedure, Eqs.\ (\ref{D-final}), (\ref{R-final}),
the coefficients of $a^2$ reduce to $\Ds d_2[0]=\frac1{3}$ 
(as for the standard BLM procedure),
the coefficients $\tilde{\tilde{d}}_3\ (\tilde{\tilde{r}}_3)$ of $a^3$, 
reduce to $0$,
while the values of the new normalization
scale $\tilde{t}_{1}~(t(\tilde{s}_{1}))$ are moderate and under control.
This effect -- to hold the control over both the types of expansions -- is 
the main advantage of
the $x$BLM\ procedure, 
while the condition $\tilde{\tilde{d}}_3 = 0$ used here 
is not a mandatory requirement for it.
Instead of that one can require to significantly reduce the size of $d_3$
for better convergence.
This trick to improve the convergence of the truncated series can be
further generalized.

\textbf{Matrix-form generalized BLM procedure}.
To generalize the seBLM\ procedure in the way mentioned above,
let us introduce a lower triangular matrix
$X= \{x_{ij}\}, ~x_{ii}\equiv 0$ associated with the matrix $Y$
instead of employing a single scalar parameter $x$.
The element $x_{ij}$ determines those portions of the contribution $y_{ij}$
which should be absorbed into the coupling renormalization,
according to the operation $x_{ij}~y_{ij}$.
On the other hand, the remainder of the contribution $y_{ij}~\bar{x}_{ij}$
(where $\bar{x}_{ij}\equiv 1-x_{ij}$) is kept in the PE-coefficient $D_i$.
The choice $X=0$ brings us back to the initial series before performing any transformations,
while $X=\{x_{21}=1,~x_{i>j}=0 \}$
corresponds to the standard BLM\ procedure.
The matrix $X=\{x_{i>j}=1 \}$ corresponds to the seBLM\ procedure, 
whereas all off-diagonal elements $y_{ij}$ should be absorbed in new couplings.
For the latter case, the first column $x_{i1}=1$ itself leads to the first stage of seBLM, as one infers by
comparing the second column in Table \ref{general-1-step} with Eq.\ (\ref{final-1-step}).
For the discussed $x$BLM procedure
the first column transforms into
$\{1, ~A_1 y_{21}\bar{x}_{i1}, ~A_1^2 y_{31}\bar{x}_{31},~\ldots\}$
instead of the diagonalized structure $\{1,~0,~0,~\ldots\}$ in seBLM. 
The schematic sketch of the first stage of $x$BLM\ 
is demonstrated in Table \ref{general-1-step}
and Eqs.\ (\ref{Delta1G0})--(\ref{Delta1G2}) and one should compare them in conjunction with the 
seBLM procedure, presented in
Table \ref{struc-2-step} (or in Eq.\ (\ref{final-1-step})).

\begin{table}[ht]
\caption{\la{general-1-step}
The structure of  $A_1$ $\cdot A_1^{i-1} y^{(1)}_{ij}(X) b_0^{-j+1}$ 
at the first stage of the $x$BLM procedure}
\centerline{
\begin{tabular}{|r|l||lll|}
\hline
                   &                      &              &        &\\
$A_1 \cdot $ &   $~1$               &              &        &\\
\hhline{|~|-|===|}
                   &                      &              &         &   \\
$A_1 \cdot $ & $(A_1)^1~y_{21}\bar{x}_{21} $& $\Ds ~+ A_1~~~\frac{y_{22}}{b_0}$ 
                   & &\\
                   &                 &              &         &\\
$A_1 \cdot $ &$(A_1)^2~y_{31}\bar{x}_{31}$& $\Ds ~+ (A_1)^2 \frac{y_{22}}{b_0}
\left(\frac{y_{32}}{y_{22}} -2 (y_{21}~x_{21}) \right) $&$\Ds ~+ (A_1)^2 \frac{y_{33}}{b_0^2} $&  \\
                   &                 &              &        &        \\
$\vphantom{^|_|}$  & $~+\ldots$        & $~+\ldots$     &$~+ \ldots$& \\
$A_1 \cdot $&$(A_1)^n~ y_{n1}\bar{x}_{n1}$& $~+\ldots$
                                                   &  $~+\ldots$ &    \\ \hline
\end{tabular}
}
\end{table}
\vspace{-5mm}

\be \la{Delta1G0}
\bar{A}_{}(t)  \to  \bar{A}_{}(t_1)\equiv A_1;
~t-t_1\equiv \Delta_{1} = \Delta_{1,0}(X) + A_{1}\cdot \Delta_{1,1}(X) + \ldots. 
\ee
The formulae for the $\Delta_1$ proper scale, Eqs.\ (\ref{Delta-0}), (\ref{Delta-1}),...
remain valid in this case by taking into account the obvious changes
$y_{i1} \to y_{i1}x_{i1}$,
\ba
\Delta_{1,0}(X)&=& y_{21} x_{21} ;\la{Delta1G1}\\
\Delta_{1,1}(X)&=& y_{31} x_{31} - 2(y_{21} x_{21})y_{21}+ (y_{21}x_{21})^2
- (y_{21}x_{21})c_1 .\la{Delta1G2}
\ea
Additional free parameters $x_{ij}$ give rise to a total amount of parameters
$\Ds n(n-1)/2$ in N$^n$LO of the PE. This allows one to perform a ``fine tuning''
of the coefficients of the series minimizing those expansion coefficients pertaining to the
considered order of the PE.
A more complicated structure of the final PE series is the price one has to pay for such
an improvement of the convergence of the series.

At the second stage of the $x$BLM procedure the contribution from the second column $y_{i2}$
transforms into the sum $\sum \Ds (A_2)^{i-1} (y_{i2}\bar{x}_{i2})/b_0$
that should be compared with the single term $(A_2)^{i-1} \delta_{i 2}~y_{22}/b_0 $
in the seBLM one.
Here $A_2$ is fixed by the modified condition for $\Delta_{2}$
\ba \la{Delta2G}
\bar{A}_{}(t_1)  \to  \bar{A}_{}(t_2)\equiv A_2;
~t_1-t_2\equiv \Delta_{2}& = &\Delta_{2,0}(X) + A_{2}\cdot \Delta_{2,1}(X) + \ldots, \nonumber \\
\Delta_{2,0}(X)&=& \frac{y_{32}}{y_{22}}x_{32} -2 (y_{21}~x_{21})
\ea
with still higher terms being omitted for simplicity and whereas the whole expression
is similar to the one in Eq.\ (\ref{2stagedelta}).
All other portions $y_{i 2}~x_{i 2}$ of this column should be absorbed into the new coupling $A_2$
by means of the expansion coefficients $\Delta_{2,i}$.
Further stages of the implementation of the $x$BLM\ procedure are similar to the first
ones and result in 
\ba \label{xBLM-final}
S(A) \to S_{x\text{BLM}} = d_0 + \frac{A(t_1)}{b_0} \cdot d_1 \left[1+
\sum_{i\geq j \geq 2}\left(\bar{A}(t_j)\right)^{i-1}\bar{x}_{i j}~y^{}_{i j} \cdot b_0^{-j+1}
\right]\, .
\ea 
In Eq.\ (\ref{xBLM-final}) we lose the clear diagonal form,
like Eq.\ (\ref{BLM-final}),
but obtain more flexibility to fix the values of the $A^i D_i$ terms in the expansion.
The elements of the X matrix for various BLM procedures, are collected in Table \ref{cases}.
\begin{table}[h]
\caption{\la{cases}
The stages of the generalization of the BLM\ procedure 
}
\begin{tabular}{|c||c|c|c|c|c|}
\hline
$\vphantom{^|_|}$ ``fine tuning'' matrix &Standard PT&BLM&seBLM (first step)&
    seBLM & $x$BLM  \\ \hhline{|~||-|-|-|-|-|}
$\vphantom{^|_|}$       $X_{ij}$              &     0     &$\delta_{i2}\delta_{1j}$&
          $\theta(i>1)~\delta_{ii}\delta_{1j}$&
          $\theta(i>j)~\delta_{ii}\delta_{jj}$&  $~\theta(i>j)~x_{ij}~ $ \\  \hline 
          &                                   &           &                        &   
          &Table~\ref{Tab:DR-3};  \\                  
$\vphantom{^|_|}$results for PT series  &Eq.\ (\ref{eq:seriesA})& Eq.\ (\ref{BLM-1})& 
Eqs.\ (\ref{final-1-step}), (\ref{Delta1})& 
              Eqs.\ (\ref{BLM-final}), (\ref{BLM-fraction})             &    \\
           &             &           &             & 
                  & Eq.\ (\ref{xBLM-final})  \\   \hline
\end{tabular}
\end{table}                    

Let us pause for a moment to make some clarifying remarks to a few partial cases:
(i) The FAC\ setting in N$^2$LO  corresponds to the condition
$\Ds (A_1)^2 \left(y_{21}\bar{x}_{21}+\frac{y_{22}}{b_0}\right)=0$ which
has been considered in Eq.\ (\ref{FAC-1}).
(ii) On the other hand, if we restrict ourselves, say,
to the N$^3$LO, 
 then
we have a matrix $X$ with 3 parameters, $x_{21},x_{31},x_{32}$ to
optimize the contributions of $A^2 D_2$ and $A^3 D_3$, respectively,
\ba
A^2 D_2 \to&&A_1\left[ A_1 y_{21}\bar{x}_{21}+
 A_2\frac{y_{22}}{b_0}  \right] = C_2; \la{gBLM-1}\\
A^3 D_3 \to&&A_1\left[ (A_1)^2 y_{31}\bar{x}_{31}+
(A_2)^2 \frac{y_{22}}{b_0}
\left(\frac{y_{32}~\bar{x}_{32} }{y_{22}}\right) +
(A_2)^2 \frac{y_{33}}{b_0^2} \right] = C_3;\la{gBLM-2}\\
\mbox{where}&& ~~~A_1= \bar{A}(t-\Delta_{1}), ~A_2=
\bar{A}(t-\Delta_{1}- \Delta_{2}) \la{gBLM-3},
\ea
and $C_2,~C_3$ are appropriate for us numbers, e.g., $C_2=C_3=0$.
The case discussed in Sec.\ \ref{Sec:GBLM} in conjunction with
Table \ref{Tab:DR-3} corresponds to the partial
solution of the above equations for $C_3=0$, $A_1=A_2$  with
$x_{21}=1$, $x_{31}=x$, $x_{32}=0$.
A complete set of solutions to Eqs.\ (\ref{gBLM-1})--(\ref{gBLM-3}) for $C_2=C_3=0$
with respect to $x_{ij}$ can be obtained and analyzed numerically.

\section{Conclusions}
In this paper we have considered the generalization of the BLM\ procedure
with the aim to 
(i) extend it sequentially 
to any fixed order of the perturbative expansion in QCD
and investigate the property of this extension, 
(ii) improve this generalization with respect to a better convergence of the perturbative
expansion.
To this end, a new hierarchy of the contributions to the expansion coefficients of
a two-point function was designed which exploits the following
properties of the beta-function coefficients: $b_i \sim b_0^{i+1}$.
Employing this new hierarchy, we constructed a sequential and unambiguous generalization 
of the BLM\ procedure \cite{BLM83} to any fixed order of the perturbative expansion
beyond the so-called ``large-$b_0$'' approximation,
a procedure we called seBLM.
This seBLM\ procedure leads to a new expansion series in terms of  
a set of new couplings $a_j=\bar{a}(t_j)$,
Eq.\ (\ref{BLM-final}), depending on a set of proper scales $t_j$ 
that can also be rewritten in a form close to a continued-fraction 
representation, given by Eq.\ (\ref{BLM-fraction}).
The advantages and disadvantages of the seBLM\ procedure were exemplarily discussed in
terms of the four--loop Adler D--function.
We found that the seBLM\ procedure may still fail in improving the perturbation-series expansion
for the D--function at its second stage,
just for the same conditions under which also the standard BLM procedure fails.
To improve the convergence of this series, the seBLM\ procedure was equipped
by additional free parameters, encoded in a parameter $x$ 
(see Eqs.\ (\ref{D-xBLM})--(\ref{D-final}) in 
Sec.\ \ref{Sec:GBLM})  
in the spirit of the Fast Apparent Convergence procedure \cite{FAC82}. 
These free parameters enable us to reduce the expansion coefficients of higher orders
in a systematic way and at the same time to control the value of the couplings $\bar{a}(t_j)$.
This improved procedure has been successfully 
applied to the series of the D-- and $\rm{R}_{(e^+e^- \to h)}$-- functions
with results presented in Table~\ref{Tab:DR-3}.
The final generalization of the BLM\ procedure is provided by the $x$BLM\ one,
which is now parameterized by a whole matrix $X=\{x_{ij}\}$ of additional parameters 
in order to achieve a better convergence
of the perturbative expansion, as it was briefly discussed in the second part of Sec.\ \ref{Sec:GBLM}.
This way, one is enabled to control both the size of the expansion coefficients and the
magnitude of the coupling $a_j$---and also of the associated scales 
$t_j$---from the most general point of view.

\acknowledgments
The author is also grateful to
A.~Grozin and N.~Stefanis for reading the manuscript and clarifying criticism,
and also to A.~Bakulev, K.~Chetyrkin, A.~Dorokhov, M.~Kalmykov,
 A.~Kataev, D.~Kazakov, A.~Kotikov, and D.~Shirkov for fruitful discussions.
This investigation has been supported by the Russian Foundation for Fundamental Research (RFBR),
 grants No.\ 05-01-00992 and No.\ 06-02-16215.

\begin{appendix}
\appendix
\section{Proof of the convolution representation}
 \label{app:A}
 \renewcommand{\theequation}{\thesection.\arabic{equation}}
  \label{App-Expansion}\setcounter{equation}{0}
\textbf{1}. Let $\bar{A} \equiv \bar{A}_{(l)}$ be a solution of the RG equation 
with the $l$--loop $B$-function.
Let us introduce the operator $\Ds \hat{D}_t = \frac{\bar{A}^2}{B(\bar{A})} \frac{d}{d t}$.
Then, we find $\hat{D}_t \bar{A} = \bar{A}^2$ and
\be \label{eq:1-loopD}
     \frac{1}{n!}\left( \hat{D}_t\right)^n~\bar{A} = \left(\bar{A} \right)^{n+1}.
\ee
Substituting (\ref{eq:1-loopD}) and Eq.\ (\ref{def:P}) into Eq.\ (\ref{eq:seriesA}),
and collecting the corresponding terms in the exponent for, e.g., $D^+(n)$ part of the series,
\ba
\sum_{n=1}D^+(n) \bar{A}^n &=& \int_0^{\infty} {\cal P}^+(\alpha)
\left(
\sum_{m=0} \frac{ \left(\alpha \hat{D}_t \right)^m}{m!}
\right)
~\bar{A}(t) d\alpha =
\int_0^{\infty} {\cal P}^+(\alpha)\left[\exp{\left(\alpha \hat{D}_t\right)} \bar{A}(t)\right] 
d \alpha \nonumber\\
&=& \int_0^{\infty} {\cal P}^+(\alpha)~ \bar{A}\left(t + \alpha \cdot
\frac{\bar{A}^2}{B(\bar{A})}\right)
 d\alpha,
\ea
one arrives at expression (\ref{Int-N-loop}).
In the two-loop case the exact solution can be expressed in
terms of the Lambert function $W(z)$, \cite{CGHJK96,Mag99} 
 defined by
\begin{eqnarray}
z=W(z)\exp\left( W(z)\right).
\end{eqnarray}
This solution has the explicit form
\begin{eqnarray}
 \label{eq:App-Exactsolution}
 \bar{A}_{(2)}(t) =
 -\frac{1}{c_1} \frac{1}{1+W_{-1}(z_t)}\, ,
 \end{eqnarray}
where $z_t = \left(1/c_1\right) \exp\left[\Ds -t_{}/c_1 -1+i\pi \right]$
and the branches of the  multivalued function $W$ are denoted by $W_{k}$, $k=0,\pm 1,\ldots $.
A review of the properties of this special function can be found in
\cite{CGHJK96}; see also \cite{Mag99}. 

\textbf{2}. In the same way, one can obtain this representation for the 
\textit{non-power} expansion series
$\{A_{n}^\text{APT}\}$ that appears instead of the standard powers $\{A^{n}\}$ 
in the analytic perturbation theory (APT) \cite{SS97,Shir99}.
The APT expansion of $S(A)$ in (\ref{eq:seriesA}) can be written as
\be \label{APT}
\Ds
S(A)=d_0 + \frac{d_1}{b_0} \cdot \sum_{n=1} D_n~A^n \to S^{\text{APT}}
=d_0 + \frac{d_1}{b_0} \cdot \sum_{n=1}D_n~A^\text{APT}_n.
\ee
The one-loop APT coupling constant is a bounded function of $t$,
 $\Ds A^\text{APT}_{1}(t)= \frac{1}{t} - \frac{1}{e^t -1}\leq 1$.
The solutions for higher values of the index $n$ of the one-loop constants $A^\text{APT}_{n}$
are not powers of $A^\text{APT}_{1}$, but they can be obtained from $A^\text{APT}_{1}$
by means of the same relation expressed in Eq.\ (\ref{1-loopRG}), see \cite{Shir99}, namely,
 \be
 A^\text{APT}_{n+1}(t) =
 \frac1{n!}\left(- \frac{d}{d t}\right)^n A^\text{APT}_{1}(t) \neq (A^\text{APT}_{1}(t))^{n+1}.
 \ee
This leads to the same convolution structure given by Eq.\ (\ref{Int-1-loop})
with $A^\text{APT}_{1}(t)$ entering into the integrand.
The finiteness of $A^\text{APT}_{1}(t)$ guarantees the convergence of this integral representation.
For this reason the corresponding series in APT, represented by Eq.\ (\ref{APT}), 
are really well-convergent \cite{fapt05}.

\textbf{3}.
The required $\beta$-function coefficients with the Minimal Supersymmetric Model (MSSM)\ light gluinos \cite{CCS97} are,
\begin{eqnarray}
 \label{eq:beta-b0}
  b_0\left(N_f, N_g\right)
   &=& \frac{11}{3} C_A - \frac{4}{3}\left( T_R N_f + \frac{N_g C_A}{2}\right)\,;\\
 \label{eq:beta-b1}
  b_1\left(N_f, N_g\right)
     &=& \frac{34}{3}C_A^2
       - \frac{20}{3}C_A \left( T_R N_f + \frac{N_g C_A}{2}\right)
      -4\left( T_R N_f C_\text{F}+ \frac{N_g C_A}{2} C_A\right);\\
 \label{eq:beta-b2}
  b_2\left(N_f, N_g\right)
     &=& \frac{2857}{54}C_A^3
       - N_f T_R \left( \frac{1415}{27}C_A^2 +\frac{205}{9}C_A C_F -2 C_F^2 \right)
       + (N_f T_R)^2 \left( \frac{44}{9}C_F +\frac{158}{27}C_A \right) - \nonumber \\
       && \frac{988}{27}N_g C_A (C_A^2) +
       N_g C_A N_f T_R \left( \frac{22}{9}C_A C_F +\frac{224}{27}C_A^2 \right)
       +(N_g C_A)^2 \frac{145}{54} C_A \, .
\end{eqnarray}
The $b_3$ coefficient, which includes the MSSM\ light gluinos, is not yet known, so
 we present it here in the standard \cite{RVL97} simplest form 
\begin{eqnarray}  
 b_3(N_f) & = &  \left( \frac{149753}{6} + 3564 \zeta_3 \right)
        - \left( \frac{1078361}{162} + \frac{6508}{27} \zeta_3 \right) N_f
   + \left( \frac{50065}{162} + \frac{6472}{81} \zeta_3 \right) N_f^2 + \nonumber \\ & &
              \frac{1093}{729}  N_f^3 \, .
\end{eqnarray} 

\section{Explicit formulae for $y^{(1)}_{i,k}$ and $\Delta_{i,k}$ }
 \renewcommand{\theequation}{\thesection.\arabic{equation}}
   \label{App-B}\setcounter{equation}{0}
The explicit expressions for the few first elements of the matrix $Y^{(1)}$ are  
\ba \la{final-1-step}
\bar{A}^1~\cdot D_1 \to&A_1^1~\cdot&~1;\nonumber\\
\bar{A}^2~\cdot D_2 \to&A_1^2~\cdot&~0 + \frac{y_{22}}{b_0}; \nonumber\\
\bar{A}^3~\cdot D_3 \to&A_1^3~\cdot&~0 +\frac{1}{b_0} \left(y_{32} -2y_{21}~y_{22} \right)
 + \frac{y_{33}}{b_0^2}; \nonumber\\
\bar{A}^4~\cdot D_4 \to&A_1^4~\cdot&~0 + \frac{1}{b_0} \left(y_{42} -3 y_{21}~y_{32}
+ y_{22}[ 5 y_{21}^2 -2 y_{31}]\right)+ \nonumber \\
&& \hspace{4.8cm} \frac{1}{b_0^2} \left(y_{43} -3y_{21}~y_{33} \right)+
\frac{y_{44} }{b_0^3} ;\nonumber\\
A^n~\cdot D_{n} \to&A_1^{n}~\cdot&~0+\frac{1}{b_0}\left(y_{n2}-\ldots \right)  \ldots ~~~~~.
\ea

The explicit expressions for the elements of the proper scales
$\Delta_1$ and $\Delta_2$ are given by
 \ba
&&\Delta_{1,0} = y_{21}=d_2[1];\\
&&\Delta_{1,1} =y_{31}-(y_{21})^2 - c_1y_{21} =\nonumber \\
&&\phantom{\Delta_{1,1}}=d_3[2]- d^2_2[1]+c_1 \left(d_3[0,1] - d_2[1]\right); \\
&&\Delta_{1,2} =y_{41}-3y_{31}y_{21}-2(y_{21})^3 -c_1 y_{31} + \frac3{2} c_1(y_{21})^2 +
(c_1^2-c_2)y_{21}=\nonumber \\
&&\phantom{\Delta_{1,1}}=d_4[3]-3d_2[1]d_3[2]+2(d_2[1])^3 + \nonumber \\
&&\phantom{\Delta_{1,1}=}c_1
\left(d_4[1,1]-3d_3[0,1]d_2[1]+\frac3{2}(d_2[1])^2 - d_3[2]\right)+ \nonumber \\
&&\phantom{\Delta_{1,1}=}c_1^2 \left(d_2[1]- d_3[0,1]\right)+
c_2 (d_4[0,0,1]-d_2[1]) \la{B-Delta(1,2)};
\ea
\ba
&&\Delta_{2,0} = \frac{y_{32}}{y_{22}}-2 y_{21} = \frac{d_3[1]}{d_2[0]}-2d_2[1];\\
&&\Delta_{2,1} = \frac{y_{42}}{y_{22}}-3 y_{21}
\frac{y_{32}}{y_{22}}+5 y_{21}^2-2y_{31}-\Delta_{2,0}^2
-c_1 \Delta_{2,0} ; \nonumber \\
 && \ldots~. \nonumber
\ea

\section{Distribution sense of \lowercase{s}BLM.}
 \renewcommand{\theequation}{\thesection.\arabic{equation}}
  \label{App-Distribution}\setcounter{equation}{0}
It is instructive to illustrate the discussed seBLM procedure
from the ``distribution''-point of view, which is based on the convolution representation provided by
Eq.\ (\ref{Int-N-loop}).
This has been demonstrated in Eq.\ (\ref{blm1}) for the one-loop running coupling explicitly.
One can re-derive the results of the seBLM\ procedure at the first stage using these terms,
e.g., for the ${\cal P}^{+}$ case.
This goes as follows.
First, one expands the factor $\bar{A}_{(l)}$
in the integrand  in Eq.\ (\ref{Int-N-loop}) in the
variable $\alpha$.
Second, following the line of argument underlying Sec.\ \ref{Sec:the first stage},
one can arrive at the same Eqs.\ (\ref{Delta-0})--(\ref{Delta-2}) for the elements.
To achieve this goal,
on should specify a partial generating function $P_1$ 
describing the distribution of the elements of the first column of the matrix $Y$,
 $\int_0^\infty P_1(\alpha)\alpha^{n-1} d\alpha \equiv\langle \alpha^{n-1} \rangle_1=y_{n1}$,
and then substitute $P_1$ in the place of ${\cal P}$ at the final stage of the procedure.
As a result of the first seBLM\ stage, i.e., Eq.\ (\ref{minor-1-step}),
the convolution $\langle \bar{A}_{(2)}(t-\ldots ) \rangle_1$ 
 reduces to
$$ A_1 \equiv
\bar{A}(t-\Delta_{1})=\int_0^\infty \delta(\alpha-\Delta_{1})\bar{A}(t-\alpha) d\alpha~.$$
Therefore, the distribution $P_1$ \textit{reduces to}
$\delta(\alpha - \Delta_{1})$, an expression resembling
the standard BLM\ result $\delta(\alpha - d_2[1])$ in Sec.\ \ref{1-loop BLM task}.
This term $\Delta_{1}$ incorporates all the ``$\alpha_s$--renormalizations''
associated with the single dressed-gluon propagator with two dressed vertices and the
connected quark fields in the skeleton diagrams in Fig.\ 3(a).
The common normalizing factor $d_1$ originates from the first tree diagrams of these skeleton
Feynman graphs.
\FIGURE[!hb]{
  
 \centerline{\includegraphics[width=0.9\textwidth,height=0.18\textheight]{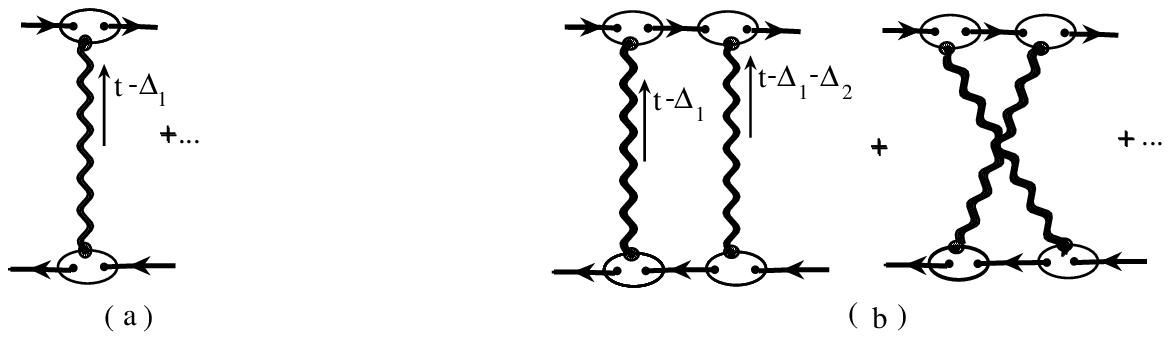}}
 \vspace*{-10mm}

   \caption{
   \footn The elements of skeleton diagrams are presented: An oval box denotes a
   dressed vertex; a thick wavy line represents a dressed gluon propagator and a
   thick fermion line stands for dressed quark propagators;
 the arrow indicates the value of the seBLM\ ``scale'' of the corresponding
 effective charge.
(a) Skeleton diagrams for the first stage of seBLM;
(b) Skeleton diagrams with two dressed gluon
    lines for the second stage of seBLM.
}
\label{F:Fig3}
}
At the second stage of the seBLM\ procedure we deal with the contribution from the
second column of $Y^{(1)}$.
Analogously, a normalized  distribution $P_2$  can be introduced for $y^{(1)}_{n2}$,
~$ \int_0^\infty P_2(\alpha)\alpha^{n-1} d\alpha
\equiv \langle \alpha^{n-1} \rangle_2= \tilde{y}_{n2}^{(1)}$.
These contributions correspond to ``two--gluon'' skeleton diagrams illustrated in Fig.\ 3(b)
(see also Fig. 2(c)).
By now the effective scale of the coupling
$\bar{A}(t-\Delta_1)$ is already fixed at the previous stage.
As a result of this operation, $P_2 \to \delta(\alpha - \Delta_{2})$ and the normalization scale
of the couplings appear to be shifted, $\bar{A}(t-\Delta_1) \to \bar{A}(t-\Delta_1-\Delta_2)$.
The normalizing factor $d_2[0]/b_0$ for that contribution originates from the undressed diagrams,
which correspond to the skeleton graphs in Fig.\ 3(b).

Executing a number of seBLM\ stages for 
$\langle \bar{A}_{(2)}(t-\ldots ) \rangle$,
 one arrives at the representation
\be
\langle \bar{A}_{(2)}(t-\ldots ) \rangle \to \int^\infty_0  d\alpha \left(
\delta(\alpha-\Delta_{1})+ \frac{d_2[0]}{b_0}
\bar{A}(t-\Delta_{1})\cdot \delta(\alpha-\Delta_{1}-\Delta_{2})
+ \ldots
\right)\bar{A}(t-\alpha).
\ee
The kernel of this convolution
can be compared with the PE representation in Eq.\ (\ref{eq:PE}),
\be
{\cal P}^{+}(\alpha) = \sum_{n} \frac{(-1)^n}{n!} \delta^{\{n\}}(\alpha) \cdot D^{+}(n) \to
\delta(\alpha-\Delta_{1})+
d_2[0]/b_0 \bar{A}(t-\Delta_{1})\cdot \delta(\alpha-\Delta_{1}-\Delta_{2})+ \ldots
\ee
The final seBLM\ series, Eq.\ (\ref{BLM-final}), corresponds to the
expansion encoded in the skeleton diagrams.
The coefficients of this expansion $d_n[0]$ ~($ y_{nn} $)
originate from the first undressed skeleton diagrams,
whereas the scales $\Delta_n$ appear due to the renormalizations in these
skeleton diagrams.

\section{Representation for the D-function}
 \renewcommand{\theequation}{\thesection.\arabic{equation}}
  \label{App-delta}\setcounter{equation}{0}
The Adler $D$ function is known \cite{Ch97} for the MSSM\ with $N_g$ light gluinos,
$D(a,N_f,N_g)$.
On the other hand, one can obtain explicit expressions for the functions
$N_f=N_f(b_0,b_1)$ and $N_g=N_g(b_0,b_1)$
solving the set of equations (\ref{eq:beta-b0}),(\ref{eq:beta-b1}) with respect to
$N_f,~N_g$. Substituting these solutions into $D(a,N_f,N_g)$, one arrives at the expansions
Eq.\ (\ref{App:d2}), (\ref{App:d3}) and Eq.\ (\ref{App:coeff-D2})--(\ref{App:coeff-D30}),
\ba
&&D(a)= 3 \sum_f Q_f^2 \left\{d_0 + d_1 \left(a
  + d_2 a^2 + d_3 a^3+ \ldots \right) \right\},~d_0 =1;~d_1=3C_F;\nonumber\\
&&d_2=b_0 d_2[1]+ d_2[0]; \la{App:d2}\\
&&d_3=b_0^2d_3[2]+b_1\cdot d_3[0,1]+ b_0 d_3[1]+ d_3[0]; \la{App:d3}\\
&&d_4=b_0^3d_4[3]+b_2\cdot d_4[0,0,1]+ b_0 b_1 d_4[1,1]+
b_0^2d_4[2]+ b_1 d_4[1]+b_0 d_4[1] +d_4[0]\, . \la{App:d4} 
\ea
The $N_f^2$--terms of $d_4$ have recently been calculated in \cite{BChK2002}, but,
unfortunately, these results cannot be used in our approach.
The reason is that it is impossible to separate the terms
$b_2 d_4[0,0,1]$ and $ b_0 b_1 d_4[1,1]$
that are of the order of $O(b_0^3)$ from the $b_0^2$--term, i.e., $b_0^2d_4[2]$,
that also contributes to the ``$N_f^2$ projection''.
This is why one needs a new degree of freedom (like the gluino).
\ba
&&d_2[1]=\frac{11}2-4\zeta_3\approx 0.691772;~~~~~~d_2[0]~~=~\frac{C_A}3-\frac{C_F}2=\frac{1}3;
\la{App:coeff-D2}\\
&&d_3[2]=\frac{302}9-\frac{76}3\zeta_3\approx 3.10345;~~~~d_3[0,1]=
\frac{101}{12}-8\zeta_3\approx -1.19979; \la{App:coeff-D32}\\
&&d_3[1]=C_A\left(\frac{3}4 + \frac{80}3\zeta_3 -\frac{40}3\zeta_5\right) -
    C_F\left(18 + 52\zeta_3 - 80\zeta_5\right)\approx 55.7005 ; \la{App:coeff-D31}\\
&&d_3[0]= \frac1{36}
    (523C_A^2 + 852 C_A C_F - 414 C_F^2) -
        72 C_A^2 \zeta_3 +
\frac{5}{24}\left(\frac{176}3-128\zeta_3\right)\frac{(\sum_f Q_f)^2}{3(\sum_f Q_f^2)} \\
&& \phantom{d_3[0]} \approx -573.9607 -19.8326\frac{(\sum_f Q_f)^2}{3(\sum_f Q_f^2)}\, .
\la{App:coeff-D30}
\ea
\end{appendix}


\begin{thebibliography}{99}
\bibitem{BLM83}
Stanley J. Brodsky and G. Peter Lepage and Paul B.
                  Mackenzie,
 {\it On the elimination of scale ambiguities in perturbative
                  {Q}uantum {C}hromodynamics},
Phys. Rev. \textbf{D28}, 228--235 (1983).

\bibitem{PDG2004}
S.~Eidelman {\it et al.},
\textit{Rev. Quantum chromodynamics},
 Phys. Lett. \textbf{B592}, 104-111 (2004).

\bibitem{FAC82}
G.~Grunberg,
\textit{ Renormalization Group Improved Perturbative QCD},
Phys.~Lett. {\bf B95}, 70 (1980), Erratum-ibid.\textbf{B110}, 501 (1982).

\bibitem{YSchr99}
Y. Schr\"{o}der,
{\it The static potential in {QCD} to two loops},
Phys. Lett.\textbf{ B447}, 321--326 (1999),
[\href{http://xxx.lanl.gov/abs/hep-ph/9812205}{{\tt hep-ph/9812205}}];
\\
 Markus Peter,
{\it The static quark-antiquark potential in QCD to three loops},
Phys. Rev. Lett. \textbf{78},602--605 (1997),
[\href{http://xxx.lanl.gov/abs/hep-ph/9610209}{{\tt hep-ph/9610209}}].

\bibitem{BrodskyL95}
     Stanley J. Brodsky and Hung Jung Lu,
{\it Commensurate scale relations in {Q}uantum {C}hromodynamics},
Phys. Rev.\textbf{D51}, 3652--3668 (1995),
[\href{http://xxx.lanl.gov/abs/hep-ph/9405218}{{\tt hep-ph/9405218}}].

\bibitem{GaKar98}
E.~Gardi and M.~Karliner,
\textit{ Relations between observables and the infrared fixed-point 
in {QCD}},
Nucl. Phys. \textbf{529}, 383--423 (1998),
[\href{http://xxx.lanl.gov/abs/hep-ph/9802218}{{\tt hep-ph/9802218}}].

\bibitem{G03}A.~G.~Grozin,
\textit{Renormalons: Technical introduction},
[\href{http://xxx.lanl.gov/abs/hep-ph/0311050}{{\tt hep-ph/0311050}}].

\bibitem{Neubert95}
M.~Neubert
\textit{ Scale setting in QCD and the momentum flow in Feynman diagrams},
 Phys. Rev. \textbf{D51}, 5924--5941 (1995),
 [\href{http://xxx.lanl.gov/abs/hep-ph/9412265}{{\tt hep-ph/9412265}}].

\bibitem{BeBr95}
M.~Beneke, V.~M.~Braun,
\textit{Naive nonAbelianization and resummation of fermion bubble
                 chains},
Phys. Lett. {\bf B348}, 513--520 (1995),
[\href{http://xxx.lanl.gov/abs/hep-ph/9411229}{{\tt hep-ph/9411229}}].

\bibitem{BaBeBr95}
Ball Patricia, M. Beneke, V. M. Braun,
\textit{Resummation of $(\beta_0 \alpha_s )^n$ corrections in QCD:
                  Techniques and applications to the tau hadronic width and
                  the heavy quark pole mass},
Nucl. Phys. \textbf{B452}, 563-625 (1995),
[\href{http://xxx.lanl.gov/abs/hep-ph/9502300}{{\tt hep-ph/9502300}}].

\bibitem{BGKL96}
Brodsky S. J., Gabadadze G. T., Kataev A. L. and Lu H. J.,
\textit{The generalized Crewther relation in QCD and its
                  experimental  consequences},
Phys. Lett. \textbf{B372}, 133--140 (1996),
[\href{http://xxx.lanl.gov/abs/hep-ph/9512367}{{\tt hep-ph/9512367}}].

\bibitem{GK92}
G.~Grunberg, A.~L.~Kataev,
\textit{On Some possible extensions of the Brodsky-Lepage-MacKenzie
                  approach beyond the next-to-leading order},
Phys.~Lett. {\bf 279}, 352--358 (1992).

\bibitem{BEGKS97} S. J. Brodsky, J. R. Ellis, E. Gardi,
                  M. Karliner and  M. A. Samuel,
\textit{Pade approximants, optimal renormalization scales, and
                  momentum flow in  Feynman diagrams},
Phys. Rev.\textbf{D56}, 6980---6992 (1997),
[\href{http://xxx.lanl.gov/abs/hep-ph/9706467}{{\tt hep-ph/9706467}}].

\bibitem{HLM02}
 Hornbostel, K. and Lepage, G. P. and Morningstar, C.,
\textit{Scale setting for $\alpha_s$ beyond leading order},
 Phys. Rev. \textbf{D67}, 034023 (2003),
[\href{http://xxx.lanl.gov/abs/hep-ph/0208224}{{\tt hep-ph/0208224}}].

\bibitem{Chyla95}
   J.~Chyla,
\textit{On the BLM scale fixing procedure, its generalizations and
                  the `genuine' higher order corrections},
Phys. Lett. \textbf{B356}, 341--348 (1995),
 [\href{http://xxx.lanl.gov/abs/hep-ph/9505408}{{\tt hep-ph/9505408}}].

\bibitem{CKS97}
 K.~G. Chetyrkin, B.~A. Kniehl, A.~Sirlin,
\textit{Estimations of order $\alpha^3_s$ and $\alpha^4_s$
    corrections to  mass-dependent observables},
   Phys. Lett. \textbf{B402}, 359--366 (1997),
   [\href{http://xxx.lanl.gov/abs/hep-ph/9703226}{{\tt hep-ph/9703226}}].

\bibitem{Mag99}
B.~A. Magradze,
\textit{ Analytic approach to perturbative QCD}, 
Int. J. Mod. Phys.
\textbf{A15},  2715--2734  (2000),
[\href{http://xxx.lanl.gov/abs/hep-ph/9911456}{{\tt hep-ph/9911456}}] 
\textit{ QCD coupling up to third order in standard and analytic perturbation 
theories},
\uppercase{D}ubna preprint E2-2000-222, 2000 (unpublished),
[\href{http://xxx.lanl.gov/abs/hep-ph/0010070}{{\tt hep-ph/0010070}}]. 

\bibitem{Mag05}
B.~A. Magradze, 
\textit{A novel series solution to the renormalization group
                  equation in QCD}, Few Body Syst.\  
\textbf{40}, 71 (2006),
 [\href{http://xxx.lanl.gov/abs/hep-ph/0512374}{{\tt hep-ph/0512374}}].

\bibitem{LL77} L.~N.~Lipatov,
\textit{Divergence of the perturbation theory series and the
                  quasiclassical theory},
 Sov.\ Phys.\ JETP 45, 216--223 (1977);\\
D.~I. Kazakov and D.~V. Shirkov,
\textit{Asymptotic series of quantum field theory and their
                  summation},
Fortsch. Phys. 28, 465--499 (1980).

\bibitem{BB00}M.~Beneke, V.~M.~Braun,
\textit{Renormalons and power corrections},
[\href{http://xxx.lanl.gov/abs/hep-ph/0010208}{{\tt hep-ph/0010208}}].

\bibitem{BGG04}V.~M.~Braun, ~E.~Gardi and ~S.~Gottwald,
\textit{ Renormalon approach to higher twist distribution 
amplitudes and the convergence of the conformal expansion},
Nucl. Phys.~\textbf{B 685},~171-226, 2004,
[\href{http://xxx.lanl.gov/abs/hep-ph/0401158}{{\tt hep-ph/0401158}}]. 

\bibitem{APiv02}~S.~Groote, J.~G.~Koerner,
A.~A.~Pivovarov, 
\textit{Spectral moments of two-point correlators in perturbation 
theory and beyond}, Phys. Rev. \textbf{D65}, 036001 (2002),
[\href{http://xxx.lanl.gov/abs/hep-ph/0105227}{{\tt hep-ph/0105227}}].

\bibitem{SS97}
D. V. Shirkov, I. L. Solovtsov,
\textit{Analytic model for the {QCD} running coupling with universal
                  ${\bar \alpha}_s(0)$ value},
 Phys. Rev. Lett. \textbf{79}, 1209--1212, 1997,
[\href{http://xxx.lanl.gov/abs/hep-ph/9704333}{{\tt hep-ph/9704333}}].

\bibitem{Shir99}
D. V. Shirkov, 
\textit{Renormalization group, causality, and nonpower perturbation
                  expansion  in {QFT}},
Theor. Math. Phys. 119, 438--447 (1999),
[\href{http://xxx.lanl.gov/abs/hep-th/9810246}{{\tt hep-th/9810246}}]. 

\bibitem{Ch97b}
K.~G.~Chetyrkin,
\textit{Correlator of the quark scalar currents and 
$\Gamma_{tot}(H \to hadrons)$  at $O(\alpha^3_s)$ in pQCD}, 
Phys.~Lett. \textbf{ B390}, 309--317 (1997),
[\href{http://xxx.lanl.gov/abs/hep-ph/9608318}{{\tt hep-ph/9608318}}].

\bibitem{GKL91}
S.~G.~Gorishny, A.~L.~Kataev,
S.~A.~Larin, 
\textit{The $O(\alpha_s^3)$ corrections to $\sigma_{tot} (e^+ e^- \to hadrons)$ 
and $\Gamma (\tau^- \to \nu_\tau + hadrons)$ in QCD},
Phys.~Lett. \textbf{B 259}, 144--150 (1991);
\\
 L.~R.~Surguladze, M.~A.~Samuel,
\textit{ Total hadronic cross-section in $e^+ e^-$ annihilation at 
the four--loop level of perturbative QCD}, 
Phys.~Rev.~Lett. \textbf{66}, 560--563 (1991).

\bibitem{Ch97}
K.~G.~Chetyrkin,
\textit{Corrections of order $\alpha^3_s$ to $R_{had}$ in pQCD with
                  light gluinos}, Phys.~Lett. \textbf{B391}, 402--412 (1997),
[\href{http://xxx.lanl.gov/abs/hep-ph/9608480}{{\tt hep-ph/9608480}}]. 
; \\
L.~J.~Clavelli and L.~R.~Surguladze,
\textit{Light gluino contribution in hadronic decays of Z boson and
                  tau lepton  to $O(\alpha^3_s)$},
Phys.~Rev.~Lett. \textbf{78}, 1632, 1997,
[\href{http://xxx.lanl.gov/abs/hep-ph/9610493}{{\tt hep-ph/9610493}}].

\bibitem{BChK2002}
P.~A.~Baikov, K.~G.~Chetyrkin, J.H.~Kuhn,
\textit{The cross section of $e^+ e^-$ annihilation into hadrons of
                  order  $\alpha^4_s~n^2_f$ in perturbative QCD},
 Phys.~Rev.~Lett. {\bf 88}, 012001, 2002,
 [\href{http://xxx.lanl.gov/abs/hep-ph/0108197}{{\tt hep-ph/0108197}}].
\bibitem{BPSS04}
     Bakulev, A. P. and Passek-Kumeri\v{c}ki, K. and Schroers, W.
                  and Stefanis, N. G.,
\textit{Pion form factor in {QCD}: From nonlocal condensates to {NLO}
                  analytic perturbation theory},
Phys. Rev. \textbf{D70} 033014 (2004),
[\href{http://xxx.lanl.gov/abs/hep-ph/0405062}{{\tt hep-ph/0405062}}].

\bibitem{CGHJK96}
R. Corless {\it et~al.},
\textit{On the Lambert W Function},
 Advances in Computation Mathematics
\textbf{5},  329  (1996). 

\bibitem{fapt05}
A.~P. Bakulev, S.~V. Mikhailov, and N.~G. Stefanis, 
\textit{QCD analytic perturbation theory: From integer powers to
                  any power of  the running coupling},
Phys. Rev. \textbf{D72}, 074014  (2005),
[\href{http://xxx.lanl.gov/abs/hep-ph/0506311}{{\tt hep-ph/0506311}}];
\textit{Fractional Analytic Perturbation Theory in Minkowski space
 and application to Higgs boson decay into a $b\bar{b}$ pair},
Phys. Rev. \textbf{D75}, 056005 (2007), 
[\href{http://xxx.lanl.gov/abs/hep-ph/0607040}{{\tt hep-ph/0607040}}].  

\bibitem{CCS97}
Clavelli, L. and Coulter, P. W. and Surguladze, L. R.,
\textit{Gluino contribution to the 3-loop beta function in the
                  minimal  supersymmetric standard model},
Phys. Rev. \textbf{D55},  4268--4272 (1997),
[\href{http://xxx.lanl.gov/abs/hep-ph/9611355}{{\tt hep-ph/9611355}}].

\bibitem{RVL97}
T. van Ritbergen, J.A.M. Vermaseren, S.A. Larin,
\textit{ The Four-loop beta function in quantum chromodynamics},
Phys.~Lett. {\bf B400}, 379--384 (1997),
[\href{http://xxx.lanl.gov/abs/hep-ph/9701390}{{\tt hep-ph/9701390}}].
\end{thebibliography}
\end{document}